\documentclass[12pt]{article}
\pdfoutput=1
\usepackage{epsfig}\parskip 5pt plus 1pt
\usepackage{bbm}
\usepackage{array}
\usepackage{hhline}
\usepackage{amsmath}
\usepackage{amssymb}
\usepackage{amsfonts}
\usepackage{mathrsfs}
\usepackage{graphicx}
\usepackage{multicol}
\usepackage{multirow}
\usepackage{fancybox}
\usepackage{slashbox}
\usepackage{pifont}
\usepackage{lscape}
\usepackage{color}
\usepackage{ulem}
\usepackage{cite}

\allowdisplaybreaks[1]

\newcommand{\nn}{\nonumber}

\newcommand{\be}{\begin{equation}}
\newcommand{\ee}{\end{equation}}
\newcommand{\bea}{\begin{eqnarray}}
\newcommand{\eea}{\end{eqnarray}}

\newcolumntype{M}[1]{>{\centering}m{#1}}

\newcommand{\equ}[1]{\eq~(\ref{equ:#1})}

\newcommand{\ie}{{\it i.e.}}

\newcommand{\eg}{{\it e.g.}}

\newcommand{\cf}{{\it cf.}}

\newcommand{\eq}{Eq.}

\newcommand{\Ref}{Ref.}
\newcommand{\Refs}{Refs.}
\newcommand{\Sec}{Sec.}

\newcommand{\App}{App.}

\newcommand{\Tab}{Table}

\begin{document}
\vspace*{-1cm}
{\begin{flushright}
FTUAM-11-40\\ IFT-UAM/CSIC-11-10\\ MPP-2011-54
\end{flushright}}
\vskip 2.0cm
\begin{center}
{\Large\bf Anomalous Higgs Couplings at the LHC, \\ and their Theoretical Interpretation}
\end{center}
\vskip 1.0  cm
\begin{center}
{\large F. Bonnet}$\,^a$~\footnote{bonnet@pd.infn.it},\, {\large M.B. Gavela}$\,^b$~\footnote{belen.gavela@uam.es},\,{\large T. Ota}$\,^c$~\footnote{toshi@mppmu.mpg.de},\,{\large W. Winter}$\,^d$~\footnote{winter@physik.uni-wuerzburg.de}\\
\vskip .2cm
$^a\,$  INFN Sezione di Padova, Via Marzolo 8, I-35131 Padova, Italy\\
\vskip .2cm
$^b\,$Departamento de F\'isica Te\'orica, Universidad Aut\'onoma de Madrid and 
\\ 
Instituto de F\'{\i}sica Te\'orica IFT-UAM/CSIC, Cantoblanco, 28049   
Madrid, Spain \\
\vskip .2cm
$^c\,$Max-Planck-Institut f\"{u}r Physik (Werner-Heisenberg-Insitut), F\"{o}hringer Ring 6, 80805 M\"{u}nchen, Germany\\
\vskip .2cm
$^d\,$Institut f\"{u}r Theoretische Physik und Astrophysik, Universit\"{a}t W\"{u}rzburg, 97074 W\"{u}rzburg, Germany 
\end{center}
\vskip 0.5cm

\begin{abstract}

We discuss  the impact and potential discovery of physics beyond the Standard Model, coupling to the Higgs sector,  at the LHC.  Using a model-independent effective Lagrangian approach, pure Higgs and Higgs-gauge operators are analyzed, and their origin in terms of  tree-level exchange of unknown heavy messengers is systematically derived.
It is demonstrated that early signals at the LHC may result from a simultaneous modification of Higgs-fermion and Higgs-gauge boson couplings induced by those operators, pointing towards singlet scalar or a triplet vector -- barring fine-tuned options. Of course, the Higgs discovery itself will also be affected by such new couplings.
With increasing statistics, the remaining options can be discriminated from each other. On the other hand, the discovery of a new scalar doublet may require technology beyond the LHC, since the Higgs self-couplings have to be measured.
   Our conclusions are based on the complete set of tree-level decompositions of the effective operators unbiased by a specific model.

\end{abstract}

\newpage
\setcounter{footnote}{0}

\section{Introduction} 
\label{sec:intro}

The Standard Model (SM) of particle physics has successfully described most of the experimental data up to now, although the abundance of free-parameters and fine-tunings  related to the origin of masses strongly suggests new physics beyond the SM (BSM). There is a plethora of specific beyond BSM theories in the literature, most of them involving new heavy fields.  In order to identify which new physics lies beyond the electroweak (EW) scale, the new parameters of such theories may be constrained by the actual, low energy, experiments. This approach requires studying each model individually, and calculating every possible observable. 

Another approach, mimicking Fermi's treatment of beta decay, consists of considering the SM as the first order approximation of the actual theory, and by completing it by a series of higher dimensional operators. When the EW symmetry breaking takes place well below the mass of the new particles, the BSM physics is taken into account -- at the EW scale and below -- by adding higher dimensional operators  to the SM Lagrangian. They are built out of SM fields and  requested to be invariant under its gauge group. Those operators are the low-energy remnant of the high energy theory. It is a beautiful and humble approach -- as it does not pretend to guess the complete high-energy model -- and it is based solely on the symmetry of the established theory. The operators are general; the model dependence is encoded in the size of the operator coefficients, which is to be set from experiments.

As stated above, it is unsatisfactory that the SM mechanism  is the sole generator of the mass of all ordinary particles (but plausibly neutrinos). If there is more to Nature than the simple unique Higgs field of the SM, it is plausible that the strength of the Higgs-matter couplings and self-couplings will depart from SM expectations. Also, even if the Higgs boson turned out to be the only particle discovered at the LHC,  the properties of the Higgs sector would be one of the major remaining construction sites, and it would be necessary to discuss the impact of possible BSM physics as model-independently as possible. 

In the literature, there exists already  substantial work on effective Higgs interactions at LEP-ILC~\cite{Hagiwara:1993qt,Grzadkowski:1995hi,Hagiwara:1996kf,Dutta:2008bh}, Tevatron~\cite{GonzalezGarcia:1999fq,Eboli:1998vg} and ILC~\cite{Barger:2003rs}. The impact of effective operators in the Higgs production at the LHC via gluon-fusion has also received some attention~\cite{Kanemura:2008ub}. Furthermore, extensive work on possible effective couplings in the context of composite Higgs models and their LHC impact has also been developed recently~\cite{Espinosa:2010vn}. On the other hand, the operator decomposition technique in terms of their possible tree-level mediator particles was developed and extensively explored in the context of non-standard neutrino interactions~\cite{Antusch:2008tz,Gavela:2008ra}. Also, the effective operators involving the Higgs field that may result from tree-level mediators have been recently studied only for the particular case of vector mediators~\cite{delAguila:2010mx}. We focus here on anomalous Higgs and Higgs-gauge effective operators analyzing: i) their independent impact on LHC signals; ii) their  decomposition in terms of tree-level mediators, which then leads to new constraints and new correlated signals.  

The first order of the effective  BSM theory consists of one
unique dimension five ($d=5$) operator that gives rise to a Majorana mass
for the neutrinos, and that it is odd under baryon minus lepton number ($B-L$) symmetry. In the following, we concentrate on $B-L$  conserving processes and  only in $d=6$ operators, expected to be the dominant ones. 
The effective Lagrangian
is then composed by the Lagrangian of the SM, $\mathcal{L}_{\mathrm{SM}}$, plus the  $d=6$ operators $\mathcal{O}_i$'s, 
\begin{equation}
\mathcal{L}_{\mathrm{eff}}=\mathcal{L}_{\mathrm{SM}}+\sum_i \alpha_i \mathcal{O}_i\,,
\label{eq:Leff}
\end{equation}
where 
 $\alpha_i$ denote the operator coefficients, which exhibit a quadratic suppression on the new physic scale (typically, the mass of the particles that have been integrated out).
Among the $d=6$ operators in which the Higgs field participates, there
 is a finite subset built from the Higgs field and the SM gauge fields
 only. We will concentrate on them in this work\footnote{%
 In Ref.~\cite{Grzadkowski:2010es}, it has been demonstrated  that the 81, $d=6$, operators of  the Buchm{\"u}ller-Wyler basis~\cite{Buchmuller:1985jz} are not all independent and should be reduced to a basis of 59 operators with the help of the equations of motion (EOM). 
In particular, it is argued that $\mathcal{O}_{\phi}^{(1)}$ is not an independent
operator from $\mathcal{O}_{\partial \phi}$ 
since it can be expressed as a combination of 
$\mathcal{O}_{\partial \phi}$
and higher dimensional Yukawa interactions   
which consist of two fermions and three Higgs doublets 
($\mathcal{O}_{e\phi}$, $\mathcal{O}_{u \phi}$, and 
$\mathcal{O}_{d \phi}$ in Ref.~\cite{Buchmuller:1985jz}),
and a $d=4$ operator $(\phi^{\dagger}\phi)^2$ in the SM.
Here we do not discuss the higher dimensional Yukawa interactions 
as the basis operators, and as a price for that,
we must treat  $\mathcal{O}_{\phi}^{(1)}$ independent 
from $\mathcal{O}_{\partial \phi}$.
}. 
In the Buchm{\"u}ller-Wyler basis~\cite{Buchmuller:1985jz}  they read:
\begin{eqnarray}
\boldsymbol{\mathcal{O}_{\phi}=-\frac{1}{3}(\phi^{\dagger}\phi)^3}\,,&& \boldsymbol{\mathcal{O}_{\partial\phi}=\frac{1}{2}\partial_{\mu}(\phi^{\dagger}\phi)\partial^{\mu}(\phi^{\dagger}\phi)}\,,\label{ophidphi}\\
\boldsymbol{\mathcal{O}_{\phi}^{(1)}=(\phi^{\dagger}\phi)(D_{\mu}\phi)^{\dagger}(D^{\mu}\phi)}\,, &&\mathcal{O}_{\phi}^{(3)}=(\phi^{\dagger}D_{\mu}\phi)((D^{\mu}\phi)^{\dagger}\phi)\,,\label{ophi1ophi3}\\
\mathcal{O}_{\phi G}=\frac{1}{2}(\phi^{\dagger}\phi)G^A_{\mu\nu}G^{A\mu\nu}\,,&&\mathcal{O}_{\phi \widetilde{G}}=(\phi^{\dagger}\phi)\widetilde{G}^A_{\mu\nu}G^{A\mu\nu}\,,\label{ula}\\
\mathcal{O}_{\phi W}=\frac{1}{2}(\phi^{\dagger}\phi)W^a_{\mu\nu}W^{a\mu\nu}\,,&&\mathcal{O}_{\phi \widetilde{W}}=(\phi^{\dagger}\phi)\widetilde{W}^a_{\mu\nu}W^{a\mu\nu}\,,\label{ulita}\\
\mathcal{O}_{\phi B}=\frac{1}{2}(\phi^{\dagger}\phi)B_{\mu\nu} B^{\mu\nu}\,,&&\mathcal{O}_{\phi \widetilde{B}}=(\phi^{\dagger}\phi)\widetilde{B}_{\mu\nu}B^{\mu\nu}\,,\label{ola}\\
\mathcal{O}_{WB}=(\phi^{\dagger}\tau^a\phi)W^a_{\mu\nu}B^{\mu\nu}\,,&&\mathcal{O}_{\widetilde{W}B}=(\phi^{\dagger}\tau^a\phi)\widetilde{W}^a_{\mu\nu}B^{\mu\nu}\,\label{olita},
\end{eqnarray}
Only the first four operators  listed, Eqs.~(\ref{ophidphi}) and
(\ref{ophi1ophi3}), can result from  tree-level exchange of heavy
particles; the rest require loop-induced
generation, or some other origin which invalidates
the expansion considered here. The size of their coefficients is
thus expected to be sub-leading 
in perturbative theories\footnote{%
However, note that in some cases operators from
new physics generated at one loop may give a large
impact on the SM loop effects. We do not discuss this possibility,
since it goes beyond the scope of our study. 
}~\cite{Arzt:1993gz,Arzt:1994gp}. 
As a  consequence, we concentrate below on the analyses of the four operators in Eqs.~(\ref{ophidphi}) and (\ref{ophi1ophi3}).

When analyzing present constraints and early LHC signals, it suffices to consider only the vacuum expectation value (vev) of the Higgs field for all operators. We will first work out the phenomenology associated to each of the four operators in Eqs.~(\ref{ophidphi}) and (\ref{ophi1ophi3}) separately,  
taking into account the constraints resulting from present electroweak precision tests (EWPT), LEP and Tevatron data. The impact on the Higgs physics at the LHC will be then discussed.
Among the four operators selected, $\mathcal{O}_{\phi}^{(3)}$ will be shown to be severely constrained beyond LHC reach, and in consequence the phenomenological analysis concentrates on the first three operators in the list above, written in bold characters.
Finally, most of the effort of this work will be dedicated to $\mathcal{O}_{\partial\phi}$ and   $\mathcal{O}_{\phi}^{(1)}$. On the other hand, $\mathcal{O}_{\phi}$  does not modify the Higgs couplings to other SM particles at tree-level, and is out of LHC reach. As far as the phenomenological part of our study is concerned, it can be regarded as generalization of \Ref~\cite{Giudice:2007fh,Espinosa:2010vn} with respect to the aspect that we study $\mathcal{O}_{\phi}^{(1)}$ as an independent operator. Therefore, we will work out the phenomenology from scratch in the $Z$-scheme, taking as inputs the
Fermi constant $G_F$ as measured from muon decay, the electromagnetic
constant $\alpha$ extracted from Thompson scattering, the $Z$ boson mass
$M_Z$ from electroweak data, and the Higgs mass $M_H$, since they are
well measured experimentally (except the Higgs mass). In addition, one of the most important new results of the phenomenology study can be found in \Sec~\ref{sec:discdev}, where the discovery of physics BSM in the effective operator framework is discussed.

As the main part of this work, we will systematically decompose each of those three operators in terms of their possible tree-level mediators and by identifying the minimal set of couplings required. This procedure allows to settle the possible SM quantum numbers of those heavy messengers, with no need of further information about the high-energy theory. It also allows to establish further constraints and new signals. Indeed, 
  the effective operator coefficients $\alpha_i$ carry information about the messengers: they will be now expressed as a combination of the high energy couplings and masses of the mediators. As a consequence, two different effective operators previously unrelated can now be linked via their effective couplings. In other words, a constraint on one of the operator coefficients  may now constrain some other coupling, even when the latter does not modify directly the low energy observables. Analogously, the new signals expected from them at LHC and elsewhere will be correlated. Finally, a separate analysis is dedicated to the theoretical implications and mediator decomposition of $\mathcal{O}_{\phi}$,  both for its intrinsic interest and eventually for future -- beyond LHC era -- use. Note that in specific models, additional signals at LHC may appear, which may come from couplings not directly related to the Higgs sector. Our work should be rather interpreted in a different direction: playing the devil's advocate, what can we learn if physics BSM shows dominantly up in the Higgs sector?
 
This paper is organized as follows: In \Sec~\ref{sec:SMLag}, the modification of the SM Lagrangian is performed in the $Z$-scheme; details of the full Lagrangian are given in \App~\ref{complete-lag}. In \Sec~\ref{section:decay}, the decay width and branching ratios of the Higgs boson are discussed in the presence of the effective interactions, where details can be found in \App~\ref{appendix:Decay}. Then in \Sec~\ref{sec:constraints}, the constraints from LEP and Tevatron are shown. 
The Higgs production at LHC is then discussed in \Sec~\ref{sec:production}. In \Sec~\ref{sec:significance}, the impact of the effective interactions on the discovery of the Higgs boson are shown, as well as the discovery of the effective operators is discussed; details can be found in \App~\ref{sec:significances}. As the next step, in \Sec~\ref{sec:theory},
  the theoretical interpretation in terms of tree level mediators is performed at the LHC, and in \Sec~\ref{sec:beyond}, perspectives for experiments beyond the LHC are pointed out. \App~\ref{App:fulldecom} gives a detailed account of the mediator decomposition. Finally, in \Sec~\ref{sec:summary} the results are summarized.


\section{Modification of the Standard Model Lagrangian}
\label{sec:SMLag}

In this section we derive the deviations, relative to the SM
predictions, induced by the effective operators considered.
Methodologically, we use the $Z$-scheme~\cite{DeRujula:1991se}
framework\footnote{
Although the radiative corrections are important in the 
electroweak measurements, we carry out our {\it renormalization} 
at the tree level, because our interest is set on the leading 
contributions of the effective operators to LHC signals.}.  The strategy of this approach can be described as follows: All SM relationships are expressed in terms of the best measured quantities $G_F$, $\alpha$, and $M_Z$, as well as $M_H$ accessible at the LHC. An effective interaction, coming from physics BSM,  will then not only show up in specific interactions directly, but also shift the SM quantities/relationships, which cannot be taken for granted anymore. Keeping the mentioned observables fixed to their measured values, we compute the impact of the direct (from the modified interaction) and indirect (from the modified SM quantities/relationships) contributions to the observables.

The  vev $v$ of the Higgs doublet $\phi$ is defined by  $\phi=(0, (v+h)/\sqrt{2})^T$, where  $h$ denotes the physical Higgs boson. As a consequence, the covariant derivative is given by  
\begin{equation}
D_{\mu}\phi=\left(\begin{array}{c} -i\frac{g_0}{2}W_{\mu}^+ (v+h) \\ \frac{1}{\sqrt{2}}\partial_{\mu}h+i\frac{\sqrt{g_0^2+g_0'^2}}{2\sqrt{2}}Z_{\mu}(v+h)\end{array}\right)\,.
\end{equation}
Substituting this expression in Eqs.~(\ref{ophidphi}) and (\ref{ophi1ophi3}) shows that three of the anomalous couplings discussed  give contributions to the kinetic energy of the Higgs boson, \ie, the Higgs field needs to be rescaled in order to get a canonical kinetic term:
\begin{equation}
\label{rescale}
h\rightarrow
(1+(\alpha_{\phi}^{(1)}+\alpha_{\phi}^{(3)}+2\alpha_{\partial\phi})\frac{v^2}{2})^{-1/2}H\,.
\end{equation}
The interaction $\mathcal{O}_{\phi}$ shifts in turn the minimum of the scalar potential, 
\begin{equation}
V(\phi)=\mu_0^2(\phi^{\dagger}\phi)+\lambda_0(\phi^{\dagger}\phi)^2+\frac{\alpha_\phi}{3}(\phi^{\dagger}\phi)^3\,,
\end{equation}
with
\begin{equation}
v^2=v_0^2(1+\alpha_{\phi}\frac{v_0^2}{4\lambda_0})\,,
\label{vev}
\end{equation}
where the subscript ``0'' denotes here and all through this paper 
the bare couplings and quantities, and in consequence 
$v_{0}^{2} \equiv -\mu_{0}^{2}/\lambda_{0}$
is the vev expression in the SM case. 
The couplings of the Higgs boson to the $Z$ and $W$ bosons turn out to include terms with high powers of the Higgs field. The complete Lagrangian at leading order in $\alpha_i$ can be found in App.~\ref{complete-lag}.

The electroweak SM contains only four independent parameters (obviating
fermion masses). We will work in the $Z$-scheme, taking as inputs the
Fermi constant $G_F$ as measured from muon decay, the electromagnetic
constant $\alpha$ extracted from Thompson scattering, the $Z$ boson mass
$M_Z$ from electroweak data, and the Higgs mass $M_H$, since they are
well measured experimentally (except the Higgs mass). At leading order,
$G_F$ and $\alpha$ are not modified by the operators in
Eqs.~(\ref{ophidphi}) and (\ref{ophi1ophi3}), while
\begin{equation}
M_Z^2=M_{Z_0}^2(1+\alpha_{\phi}^{(1)}\frac{v^2}{2}+\alpha_{\phi}^{(3)}\frac{v^2}{2}+\alpha_{\phi}\frac{v^2}{4\lambda_0})
\end{equation}
and 
\begin{equation}
M_H^2=M_{H_0}^2(1-\alpha_{\phi}^{(1)}\frac{v^2}{2}-\alpha_{\phi}^{(3)}\frac{v^2}{2}-
\alpha_{\partial\phi}v^2+3\alpha_{\phi}\frac{v^2}{4\lambda_{0}})\,.
\label{MHiggs}
\end{equation}
After renormalization, the relevant gauge and gauge-Higgs term of the Lagrangian read
\begin{eqnarray}
\mathcal{L}_{H,Z,W}&\ni&M_{W}^2 W_{\mu}^-W^{+\mu}+
\frac{1}{2} M_{Z}^2 Z_{\mu}
Z^{\mu}+\frac{1}{2}\partial_{\mu}H\partial^{\mu}H-
\frac{1}{2} M_{H}^2H^2\nn\\
&+&\lambda_{HWW}W_{\mu}^-W^{+\mu}H+\lambda_{HZZ} Z_{\mu}Z^{\mu}H-\lambda_{HHH}H^3\nn\\
&+&\lambda_{HHWW}W_{\mu}^-W^{+\mu}H^2+\lambda_{HHZZ}Z_{\mu}Z^{\mu}H^2-\lambda_{HHHH}H^4
\label{EffLag}
\end{eqnarray}
with
\begin{eqnarray}
M_W^2&=&M_{W_{\mathrm{SM}}}^2(1-\frac{c^2}{c^2-s^2}\alpha_{\phi}^{(3)}\frac{v^2}{2})\,,\label{MW}\\
\lambda_{HWW}&=&\lambda_{HWW_{\mathrm{SM}}}(1+\alpha_{\phi}^{(1)}\frac{v^2}{2}-(\frac{1}{2}+\frac{c^2}{c^2-s^2})\alpha_{\phi}^{(3)}\frac{v^2}{2}-\alpha_{\partial\phi}\frac{v^2}{2})\,,\label{HWW}\\
\lambda_{HZZ}&=&\lambda_{HZZ_{\mathrm{SM}}}(1+\alpha_{\phi}^{(1)}\frac{v^2}{2}+\alpha_{\phi}^{(3)}\frac{v^2}{4}-\alpha_{\partial\phi}\frac{v^2}{2})\,,\label{HZZ}\\
\lambda_{HHWW}&=&\lambda_{HHWW_{\mathrm{SM}}}(1+\frac{5}{2}\alpha_{\phi}^{(1)}v^2-(1+\frac{c^2}{c^2-s^2})\alpha_{\phi}^{(3)}\frac{v^2}{2}-\alpha_{\partial\phi}v^2)\,,\\
\lambda_{HHZZ}&=&\lambda_{HHZZ_{\mathrm{SM}}}(1+\frac{5}{2}\alpha_{\phi}^{(1)}v^2+2\alpha_{\phi}^{(3)}v^2-\alpha_{\partial\phi}v^2)\,,\\
\lambda_{HHH}&=&\lambda_{HHH_{\mathrm{SM}}}(1-\alpha_{\phi}^{(3)}\frac{v^2}{4}-\alpha_{\partial\phi}\frac{v^2}{2}+\frac{1}{3}\alpha_{\phi}\frac{v^2}{\lambda_{0}})\,,\label{HHH}\\
\lambda_{HHHH}&=&\lambda_{HHHH_{\mathrm{SM}}}(1-\alpha_{\phi}^{(3)}\frac{v^2}{2}-\alpha_{\partial\phi}v^2+2\alpha_{\phi}\frac{v^2}{\lambda_{0}})\, . \label{HHHH}
\end{eqnarray}
Here the subscript ``$\mathrm{SM}$'' denotes the SM prediction for the corresponding mass or coupling\footnote{Their expression in terms of the chosen observables can be found in \App \ref{complete-lag}. },  and  $c$ and $s$ denote
the cosine and sine of Weinberg angle as functions of the input parameters
\begin{eqnarray}
c^{2} \equiv \cos^{2} \theta_{W} 
&=& \frac{1}{2} ( 1 + (1- \frac{4\pi \alpha}{\sqrt{2} G_{F} M_{Z}^{2}})^{-1/2}
) \, ,
\\
s^{2} \equiv \sin^{2} \theta_{W} 
&=& \frac{1}{2} ( 1 - (1- \frac{4\pi \alpha}{\sqrt{2} G_{F} M_{Z}^{2}})^{-1/2}
) \, .
\end{eqnarray}
Because of the rescaling of the Higgs field, the Higgs-fermion couplings, 
 \begin{equation}
\mathcal{L}_{f}\ni 
\frac{Y_{f}v}{\sqrt{2}}\overline{f}f
+\frac{Y_{f}}{\sqrt{2}} H\overline{f}f\,,
\end{equation}
which induce  fermion masses $m_{f}\equiv Y_{f}{v}/{\sqrt{2}}$,  
are  also modified:
\begin{eqnarray}
\label{fermion}
\lambda_{Hff} =
\frac{Y_{f}}{\sqrt{2}}
=
\frac{Y_{f_{\mathrm{SM}}}}{\sqrt{2}}(1-\alpha_{\phi}^{(3)}\frac{v^2}{4}-
\alpha_{\partial\phi}\frac{v^2}{2})\,.
\end{eqnarray}

The operator $\mathcal{O}_{\phi}^{(3)}$ violates the custodial 
symmetry and 
contributes differently to the $W$ and $Z$ masses and 
couplings, as can be seen in the equations above~\cite{Grinstein:1991cd,Barbieri:2004qk}. It is thus very 
constrained~\cite{Nakamura:2010zzi} by present data on electroweak precision tests, such as the 
$\rho$ parameter \begin{equation}
\rho \equiv \frac{M_Z^2 c^2}{M_W^2}=\frac{M_{W_{\mathrm{SM}}}^2}{M_W^2},
\end{equation}
with, in the present case, 
\begin{equation}
\delta\rho=\frac{c^2}{c^2-s^2}\alpha_{\phi}^{(3)}\frac{v^2}{2}\,.
\end{equation}
The constraint on $\rho$ is thus tantamount to a constraint on $\alpha_{\phi}^{(3)}$. It is common to 
replace $\delta\rho$  by the $T$ parameter~\cite{Peskin:1991sw}, $\delta\rho\equiv \alpha T$.
The latest measurement~\cite{Nakamura:2010zzi} imposes $T=-0.03\pm0.11$, indicating
\begin{equation}
 \alpha_{\phi}^{(3)} v^2\lesssim 3\cdot 10^{-4}
\end{equation}
 and thus out of LHC reach. In consequence, we will disregard it for the phenomenology analysis, and concentrate below exclusively on the operators $\mathcal{O}_{\phi}$, 
$\mathcal{O}_{\phi}^{(1)}$, and $\mathcal{O}_{\partial\phi}$ in Eqs.~(\ref{ophidphi}) and (\ref{ophi1ophi3}).
 As can be seen from Eqs.~(\ref{MW})-(\ref{HHHH}), $\mathcal{O}_{\phi}$ only modifies the trilinear and quartic couplings of the Higgs boson. Such couplings will be very hard to observe at the LHC, see, \eg, \Refs~\cite{Barger:2003rs,Kanemura:2008ub,Baur:2002qd,Plehn:2005nk,Baur:2009uw}, and are not involved in the discovery searches of the Higgs boson. Therefore, $\alpha_{\phi}$ will be discussed separately in \Sec~\ref{sec:beyond}.

 As the Higgs couplings to the gauge bosons are modified, the exact
 cancellation, via the exchange of a Higgs boson, of the terms growing
 with the energy  in the longitudinal gauge bosons scattering amplitudes
 do no longer occur. Indeed, it is easy to see that  with the presence
 of the effective operators the divergent part in the high energy regime
 goes like $(\alpha_{\phi}^{(1)}-\alpha_{\partial\phi})v^2 s$ or
 $(\alpha_{\phi}^{(1)}-\alpha_{\partial\phi})v^2 (s+t)$, depending on
 the process. As a consequence, tree-level unitarity is violated at high
 enough energies.  However, this growth with energy is only valid up to
 the effective theory cut-off scale. Above that scale the fate of
 unitarity depends  on the detail of the UV completion.


\section{Higgs Branching Ratios and Decay Widths\label{section:decay}}
\begin{figure}[tp]
\begin{center}
\epsfig{figure=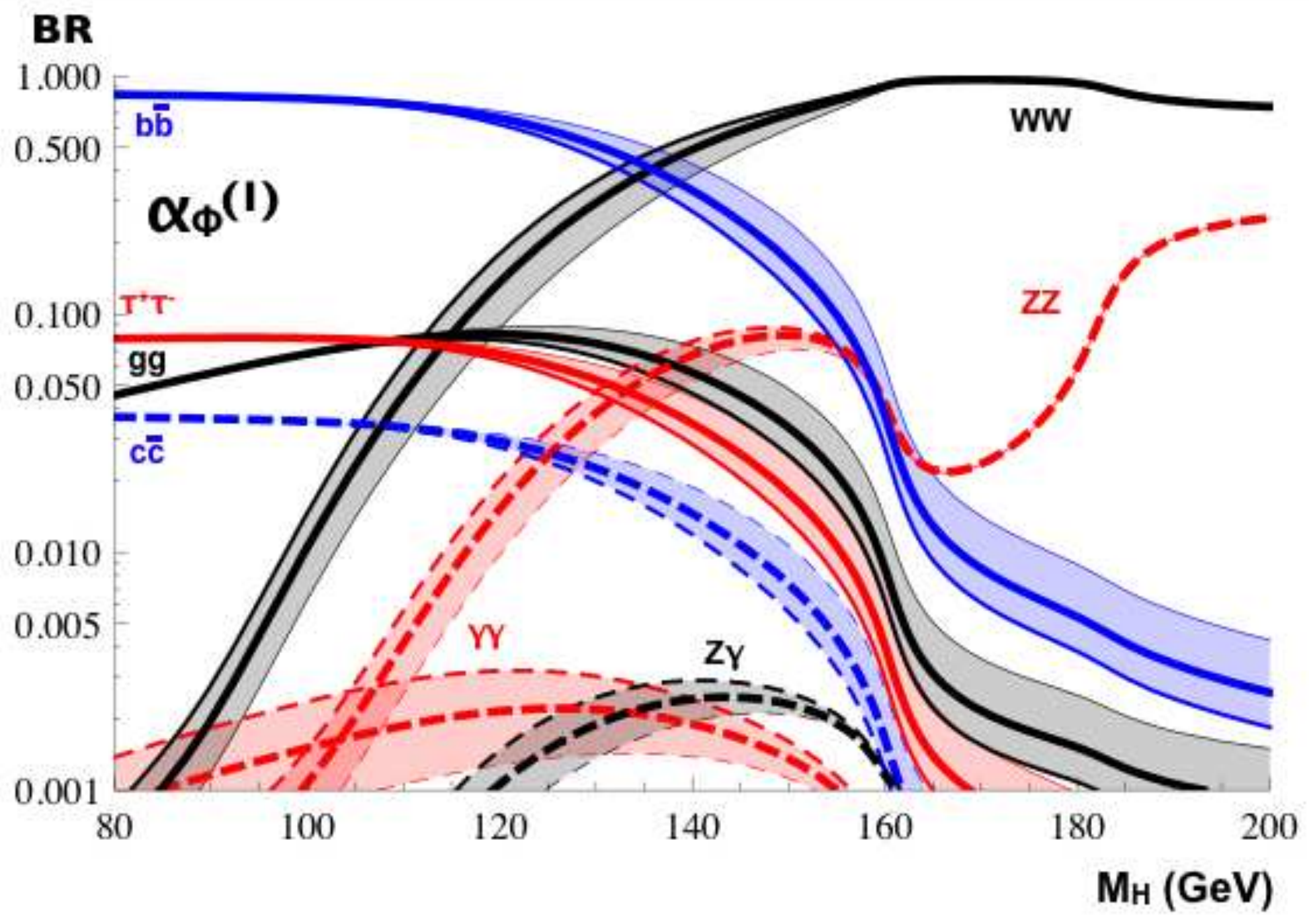,width=12cm}\\
\caption{\label{BR_xf1}
$\mathcal{O}_{\phi}^{(1)}$ impact on Higgs branching ratios, as a function of $M_H$. The thick (middle) curves represent the SM reference, and the shaded regions mark the range $-0.4 \le \alpha_{\phi}^{(1)}v^2 \le 0.4$  (thin curves for the case  $\alpha_{\phi}^{(1)}v^2=-0.4$ and medium thick curves for $\alpha_{\phi}^{(1)}v^2=0.4$).}
\end{center}
\end{figure}

Consider the impact of the effective interactions on Higgs branching ratios and decay widths for $\alpha_i\ll1$ (see Ref.~\cite{Djouadi:2005gi} and references therein). Eq.~(\ref{MW}) illustrates that the $W$-boson mass, 
which is a prediction in the $Z$-scheme, does not get modified by any of the three operators considered. 
For the different Higgs decay channels, we obtain:
\begin{description}
\item{$\boldsymbol{H\rightarrow f\overline{f}:}$}
 From Eq.~(\ref{fermion}), it follows that
\begin{equation}
\label{BRfermion}
\Gamma(H\rightarrow f\overline{f})=(1-\alpha_{\partial\phi}v^2) \, \Gamma_{\mathrm{SM}}(H\rightarrow f\overline{f})\,.
\end{equation}
\item{$\boldsymbol{H\rightarrow gg:}$}
This decay is mediated by heavy quarks loops, resulting in 
\begin{equation}
\label{BRgluon}
\Gamma(H\rightarrow gg)=(1-\alpha_{\partial\phi}v^2) \, \Gamma_{\mathrm{SM}}(H\rightarrow gg)\,.
\end{equation}
\item{$\boldsymbol{H\rightarrow VV:}$}
The modification of the vertices in Eq.~(\ref{HZW}) leads to 
\begin{eqnarray}
\label{BRVV}
\Gamma(H\rightarrow ZZ)&=&(1+\alpha_{\phi}^{(1)}v^2-\alpha_{\partial\phi}v^2) \, \Gamma_{\mathrm{SM}}(H\rightarrow ZZ)\,, \label{HZZm} \\
\Gamma(H\rightarrow WW)&=&(1+\alpha_{\phi}^{(1)}v^2-\alpha_{\partial\phi}v^2) \, \Gamma_{\mathrm{SM}}(H\rightarrow WW)\,. \label{HWWm} 
\end{eqnarray}
\item{$\boldsymbol{H\rightarrow \gamma\gamma:}$}
The SM Higgs decay into two photons is mediated  by fermion (mainly top quark)
	   and $W$ loops, and the new physics corrections are given by
\begin{eqnarray}
\label{BRphoton}
\frac{\Gamma(H\rightarrow \gamma\gamma)}{\Gamma_{\mathrm{SM}}(H\rightarrow\gamma\gamma)}=\frac{\left|(1-\alpha_{\partial\phi}\frac{v^2}{2})\frac{4}{3}A^{H}_{1/2}(\tau_t)+(1+\alpha_{\phi}^{(1)}\frac{v^2}{2}-\alpha_{\partial\phi}\frac{v^2}{2})A_1^H(\tau_W)\right|^2}{\left|\frac{4}{3}A^{H}_{1/2}(\tau_t)+A_1^H(\tau_W)\right|^2}\,,
\end{eqnarray}
where $A^{H}_{1/2}$ and $A_1^H$ are functions that can be found in App.~\ref{appendix:Decay}.
\item{$\boldsymbol{H\rightarrow Z\gamma:}$}
Again, this process it is mediated by fermions 
and $W$-boson loops, leading to 
\begin{eqnarray}
\label{BRZphoton}
\frac{\Gamma(H\rightarrow Z\gamma)}{\Gamma_{\mathrm{SM}}(H\rightarrow Z\gamma)}=\frac{\left|(1-\alpha_{\partial\phi}\frac{v^2}{2})\sum_f N_c\frac{Q_f v_f}{c}B^{H}_{1/2}(\tau_f)+(1+\alpha_{\phi}^{(1)}\frac{v^2}{2}-\alpha_{\partial\phi}\frac{v^2}{2})B_1^H(\tau_W)\right|^2}{\left|\sum_f N_c\frac{Q_f v_f}{c}B^{H}_{1/2}(\tau_f)+B_1^H(\tau_W)\right|^2}\,,\nn\\
\end{eqnarray}
where the functions $B^{H}_{1/2}$ and $B_1^H$ and the SM rate $\Gamma_{\mathrm{SM}}(H\rightarrow Z\gamma)$ can be found in App.~\ref{appendix:Decay}.
\end{description}
The   branching ratios for the Higgs decay have been computed using the  HDECAY  program~\cite{Djouadi:1997yw}, that we modified in order to take into account the effective interactions discussed in this work.\footnote{HDECAY includes most of the higher order corrections to the Higgs decays as well as off-shell effects for the Higgs decay into a pair of massive gauge bosons or a pair of top quarks.} 
\begin{figure}[tp]
\begin{center}
\epsfig{figure=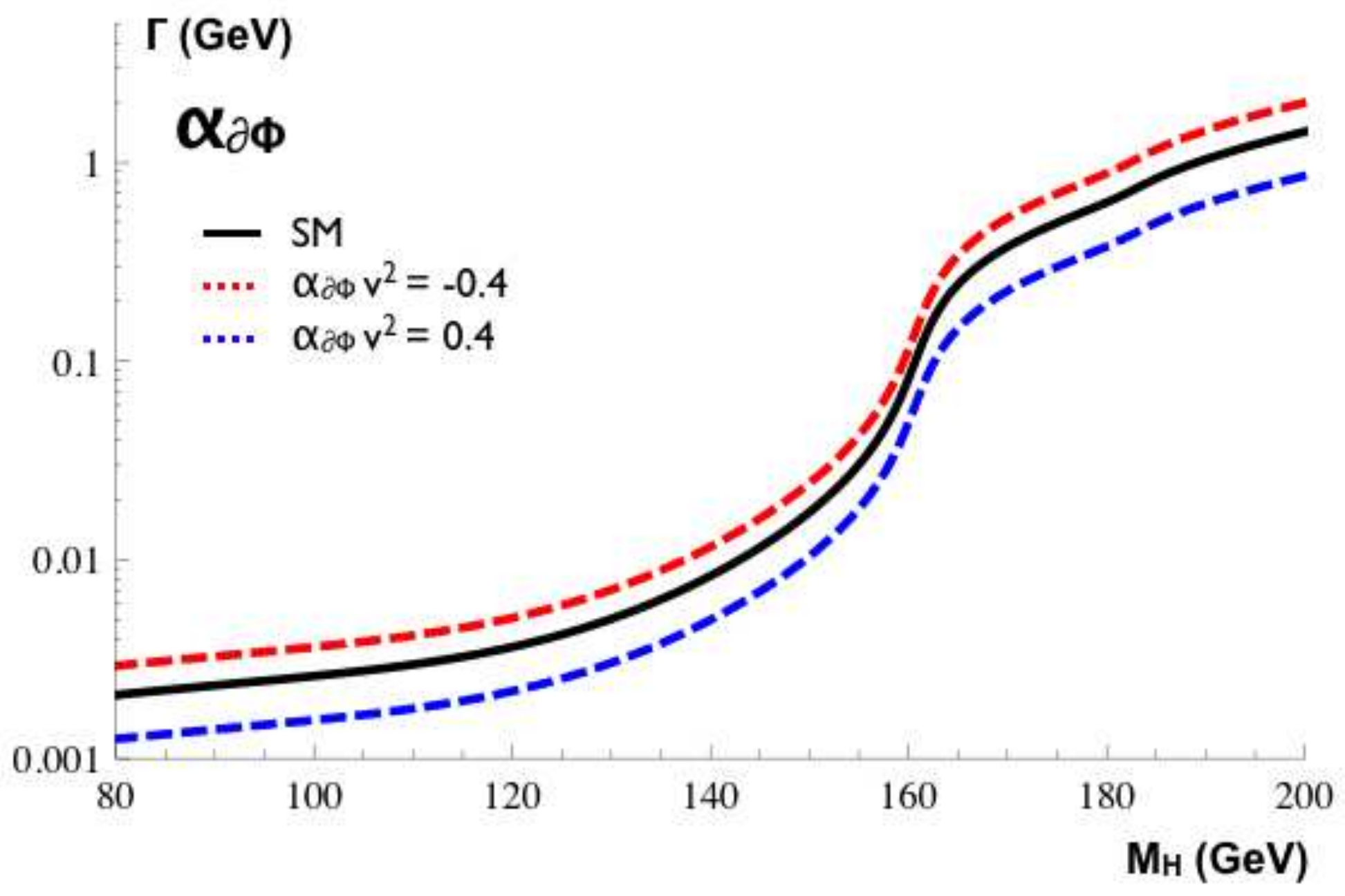,width=7.3cm}
\hspace*{0.5cm}
\epsfig{figure=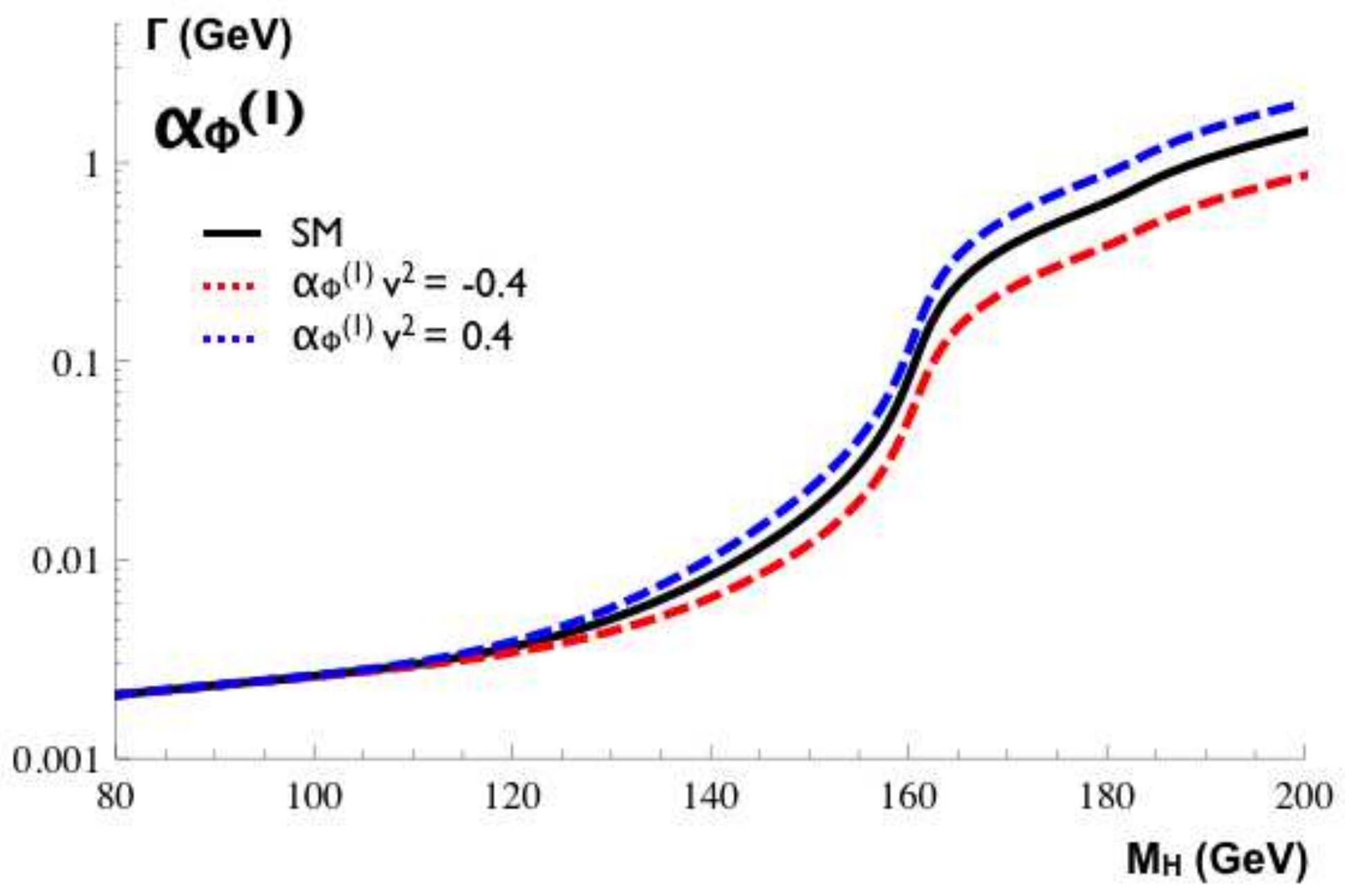,width=7.3cm}
\caption{\label{Widths}
Total decay width of the Higgs boson as a function of $M_H$ for $\alpha_{\partial\phi}$ (left) and $\alpha_{\phi}^{(1)}$ (right), for the values given in the plot legends.}
\end{center}
\end{figure}
The (total) decay width of the Higgs is shown in Fig.~\ref{Widths} as a function of  $M_H$. As shown in the left panel,  it varies linearly with $\alpha_{\partial\phi}$ independent of $M_H$ since all decay widths are modified in the same way. As a consequence,  the coupling $\mathcal{O}_{\partial\phi}$ does not affect the Higgs branching ratios, which remain equal to the SM ones.
On the other hand, for non-vanishing $\alpha_{\phi}^{(1)}$ (right panel), the decay width is modified only for large Higgs masses, where the decays into massive gauge bosons are dominant, \cf, Eqs.~(\ref{HZZm}) and~(\ref{HWWm}): we show the corresponding branching ratios in Fig.~\ref{BR_xf1} as a function of $M_H$.
 Here the thick (middle) curves represent the SM reference, and the shaded regions mark the range $-0.4 \le \alpha_{\phi}^{(1)}v^2 \le 0.4$. For large $M_H$, the decays into vector bosons clearly dominate, which means that their relative contribution does not change. However, the relative contributions from the other channels are anti-correlated with $\alpha_{\phi}^{(1)}$ because the total decay width increases while the individual channels remain unaffected. For small $M_H$, the decays into vector bosons are sub-dominant, which means that the total width is hardly affected by $\alpha_{\phi}^{(1)}$, just as the leading channels. However,  the relative contributions of the decays into vector bosons are proportional to $\alpha_{\phi}^{(1)}$. Note that also the branching ratios into photons depend somewhat on $\alpha_{\phi}^{(1)}$.  


\section{Constraints from LEP and Tevatron}
\label{sec:constraints}

\begin{figure}[tp]
\begin{center}
\epsfig{figure=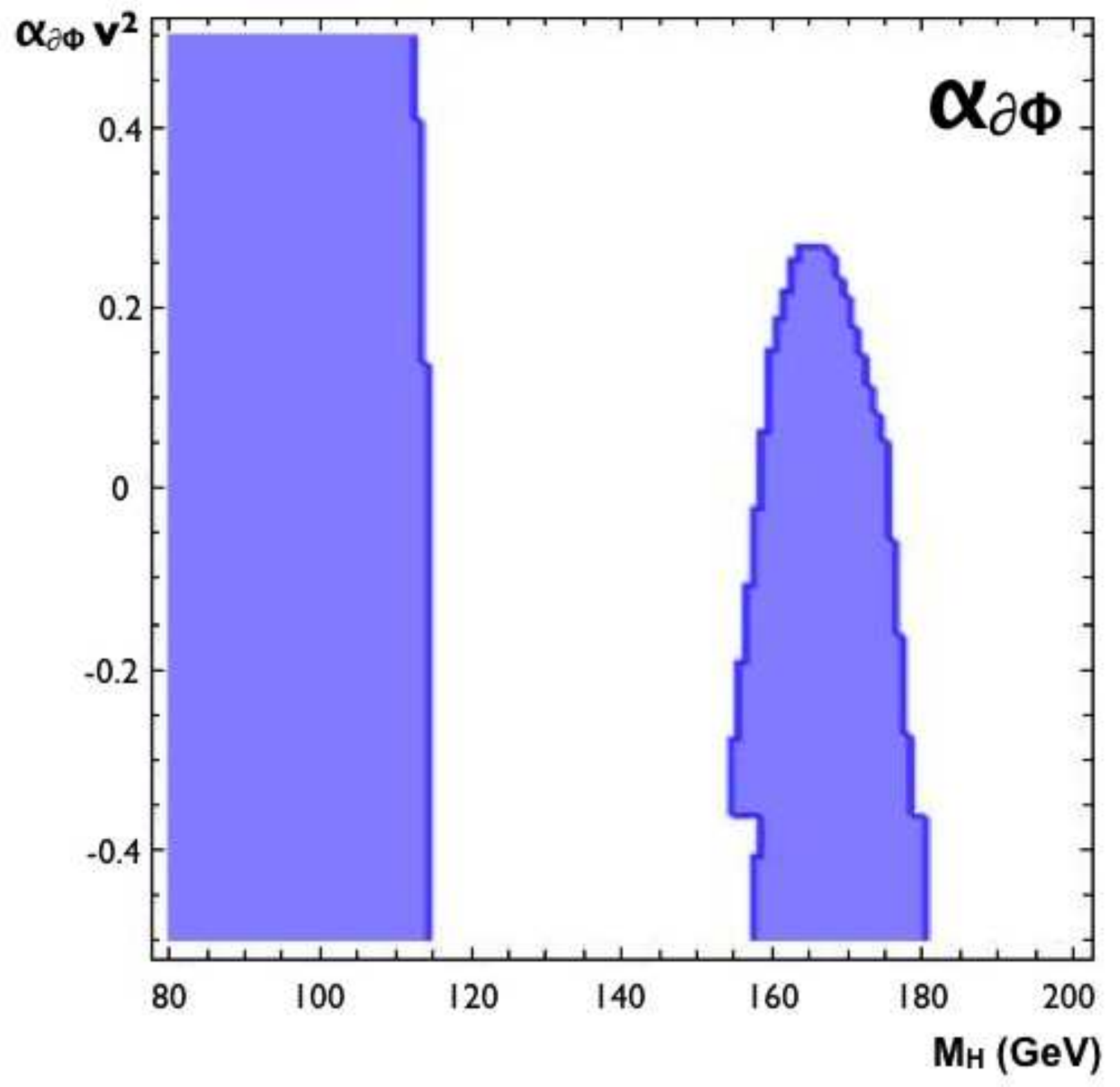,width=7.3cm}
\hspace*{0.5cm}
\epsfig{figure=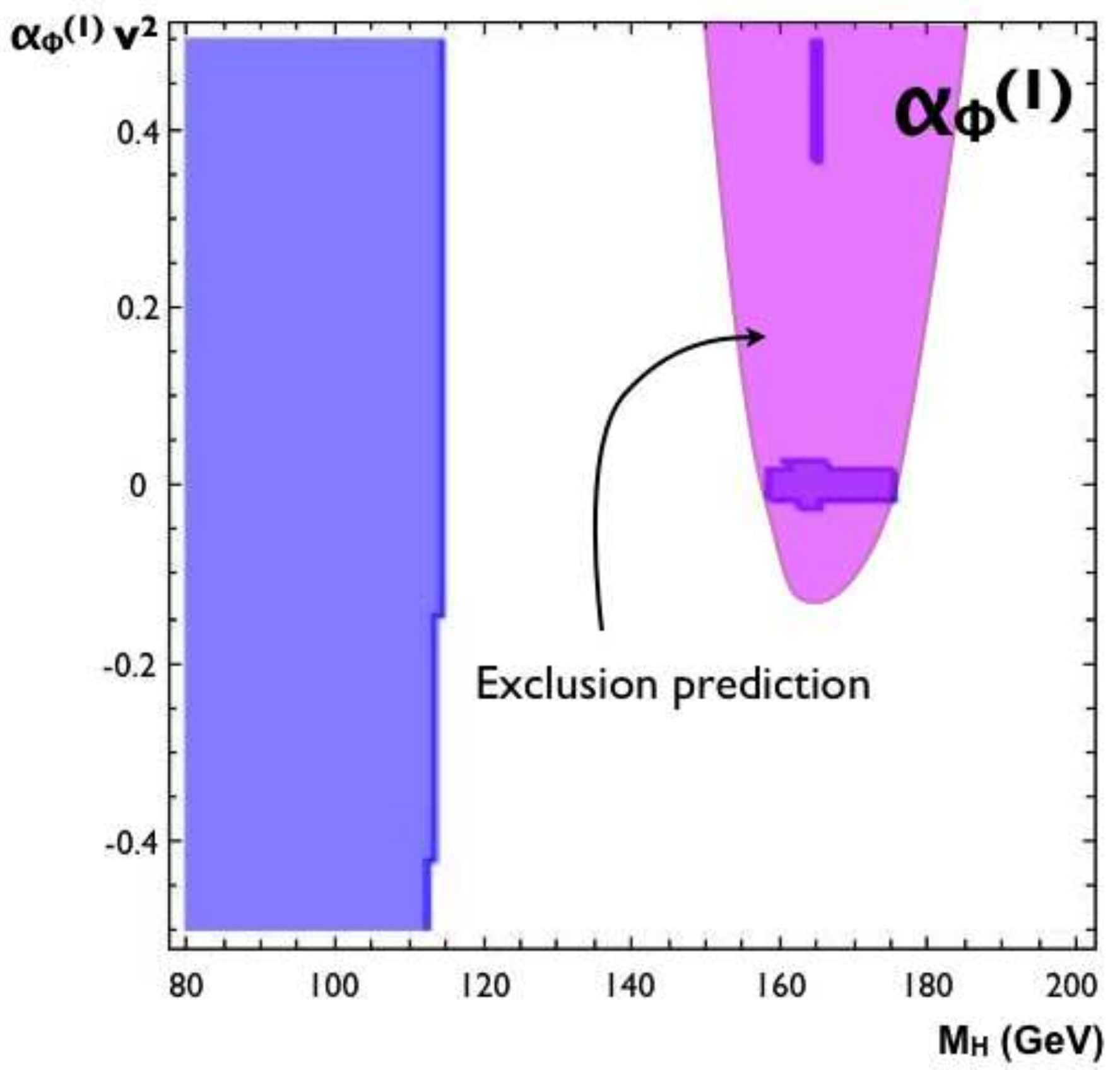,width=7.3cm}
\caption{\label{LEPTEV}Combined LEP and Tevatron experimental 95\%C.L. exclusion regions  in the ($M_H$,$\alpha_{\partial\phi}v^2$) (left) and ($M_H$,$\alpha_{\phi}^{(1)}v^2$) planes (right), obtained with the program HiggsBounds~\cite{Bechtle:2008jh,Bechtle:2011sb}. The purple region (right) indicates our prediction for the exclusion region; see main text for explanations.}
\end{center}
\end{figure}

LEP and Tevatron already set bounds on the allowed SM Higgs mass region, which may be modified in the presence of either $\mathcal{O}_{\phi}^{(1)}$ or $\mathcal{O}_{\partial\phi}$. In order to test this impact, we have modified the HiggsBounds program~\cite{Bechtle:2008jh,Bechtle:2011sb}.  

The most relevant channel at LEP is the $e^+e^-\rightarrow ZH\rightarrow
Zb\overline{b}$ search~\cite{Schael:2006cr}. This channel is sensitive
to the $\lambda_{HZZ}$ and $\lambda_{Hff}$ couplings, see
Eqs.~(\ref{HZZ}) and (\ref{fermion}).  At the Tevatron, the most sensitive channel for Higgs searches is  $H\rightarrow WW$ decay. The combined CDF and D0 analyses~\cite{CDF:2010ar} that studied this channel included Higgs production via gluon fusion, vector boson fusion, and Higgsstrahlung. For the SM, the mass range $158~\mathrm{GeV}< M_{H} <175~\mathrm{GeV}$ has been excluded at the $95\%$ CL. 
The HiggsBounds program uses the SM combined analysis only if the relative contribution to the event rate of different search channels, included in such studies,  is the same as in the SM. In other words, the Higgs boson predicted by the model tested should behave as a SM Higgs boson.
 For instance, if gluon and vector fusion  were the only relevant production modes, the condition to use SM combined limits would be
\begin{equation}
\frac{\sigma_{\text{BSM-model}}(gg\rightarrow H\rightarrow WW)}{\sigma_{\mathrm{SM}}(gg\rightarrow H\rightarrow WW)}=\frac{\sigma_{\text{BSM-model}}(WW/ZZ\rightarrow H\rightarrow WW)}{\sigma_{\mathrm{SM}}(WW/ZZ\rightarrow H\rightarrow WW)}=\mathrm{const.} \, 
\label{constant}
\end{equation}
This condition is only fulfilled if $\alpha_{\phi}^{(1)}$ is vanishing, since a non-vanishing $\alpha_{\phi}^{(1)}$ modifies only the $HWW$ and $HZZ$ vertices and not the $Hgg$ one. Thus only single channel studies can be used to constraint $\alpha_{\phi}^{(1)}$ when $|\alpha_{\phi}^{(1)}|v^2>0.02$.

We have analyzed the excluded regions in the
 parameter spaces defined by ($M_H$,$\alpha_{\partial\phi}v^2$) and ($M_H$,$\alpha_{\phi}^{(1)}v^2$) in Fig.~\ref{LEPTEV} in the left and right panels, respectively.
  As far as the ($M_H$,$\alpha_{\partial\phi}$) analysis is concerned (left plot),
both the $HZZ$ and $Hff$ couplings decrease for $\alpha_{\partial\phi}>0$, explaining the small degradation of the LEP bounds  in that region. With respect to the Tevatron bound, the  $H\rightarrow WW$ branching ratio equals that of the SM  (see Sec.~\ref{section:decay}), but the production cross section decreases (increases) for $\alpha_{\partial\phi} > 0$ ($<0$), softening (enlarging) the $M_H$ exclusion region.\footnote{The gluon fusion process is dominant but all production modes, being vector boson fusion of Higgsstrahlung, are modified in the same way and were included in the analysis. The kink of the excluded band at $\alpha_{\partial\phi}v^2\sim-0.37$ is a technical byproduct of the use of two different studies above~\cite{Aaltonen:2010sv} and below~\cite{Aaltonen:2010yv} this point.} Negative values of  $\alpha_{\partial\phi}$ are excluded in this range of $M_H$.
Note that  all vertices at stake are rescaled by the same coefficient $1-\alpha_{\partial\phi}v^2$, and the condition in Eq.~(\ref{constant}) holds.

For the ($M_H$,$\alpha_{\phi}^{(1)}$) analysis -- see right panel in Fig.~\ref{LEPTEV} -- 
 the Higgs-fermion couplings are not modified and the Higgs-gauge couplings increase with $\alpha_{\phi}^{(1)}$, which means that the LEP bounds soften for negative values of $\alpha_{\phi}^{(1)}$. At the Tevatron, the dominant production cross section $gg\rightarrow H$ is not modified, but the vector boson fusion mechanism and the decay $H\rightarrow WW$ rate get enhanced for  positive values of $\alpha_{\phi}^{(1)}$. Therefore, one would expect a sizeable  excluded region.  This is barely seen with the present available studies, as the condition to use combined analyses described in Eq.~(\ref{constant}) does not apply unless $|\alpha_{\phi}^{(1)}|v^2\leq0.02$. We expect that dedicated studies by the experimentalists of Tevatron would be able to exclude in this case a broader region of the parameter space, as we have tentatively shown in Fig.~\ref{LEPTEV} (``Exclusion prediction''). 

To summarize, while the LEP bounds for the Higgs mass are relatively robust with respect to $\mathcal{O}_{\phi}^{(1)}$ and $\mathcal{O}_{\partial\phi}$, the Tevatron bound does not hold in the presence of new physics in the Higgs sector. For instance, if $\alpha_\phi^{(1)}v^2=-0.2$, the bound disappears. In addition, a contribution of $\mathcal{O}_{\phi}^{(1)}$ and $\mathcal{O}_{\partial\phi}$ cannot be excluded from LEP and Tevatron, unless $\alpha_{\partial\phi} \lesssim 0.2$ or $\alpha_{\phi}^{(1)} \gtrsim -0.2$ in the $M_H$ range probed by the Tevatron.


\section{Higgs Production at the LHC}
\label{sec:production}

Here we summarize the modification of the Higgs production channels at the LHC:
\begin{description}
\item{{\bf Gluon Fusion:}}
 The most important Higgs boson production channel at LHC is the gluon fusion process $gg\rightarrow H$, taking place at leading order  through fermion loops (mainly bottom and top quarks). Since NLO QCD corrections do not affect the Higgs couplings~\cite{Djouadi:2005gi}, the production cross section is simply given by
\begin{equation}
\sigma_{\mathrm{NLO}}(gg\rightarrow H)=(1-\alpha_{\partial\phi}v^2) \, \sigma^{\mathrm{SM}}_{\mathrm{NLO}}(gg\rightarrow H)\,.
\end{equation}
The NLO QCD cross section was obtained with the program HIGLU~\cite{Spira:1995mt}.\\

\item{{\bf Vector boson fusion:}}
Vector boson fusion, $qq\rightarrow qq+W^*W^*(Z^*Z^*)\rightarrow Hqq$ is the second most important production mode. As for gluon fusion, the NLO QCD corrections do not depend on the Higgs boson couplings  \cite{Djouadi:2005gi} and thus 
\begin{equation}
\sigma_{\mathrm{NLO}}(VV\rightarrow H)= (1+\alpha_{\phi}^{(1)}v^2-\alpha_{\partial\phi}v^2) \, \sigma_{\mathrm{NLO}}^{\mathrm{SM}}(VV\rightarrow H)
\end{equation}
with $V=W,Z$. The NLO QCD cross section has been obtained with the program VV2H~\cite{Spiraurl}.    \\


\item{{\bf Associated production (Higgsstrahlung):}}
The radiation of a Higgs boson off a gauge boson, $q\overline{q}\rightarrow W^*(Z^*)\rightarrow W(Z)+H$ is an important production mode in the intermediate mass region. Once again
\begin{equation}
\sigma_{\mathrm{NLO}}(VH)= (1+\alpha_{\phi}^{(1)}v^2-\alpha_{\partial\phi}v^2) \, \sigma_{\mathrm{NLO}}^{\mathrm{SM}}(VH)\,.
\end{equation}
The NLO QCD cross section has been calculated with the program V2HV~\cite{Spiraurl}.\\


\item{{\bf Radiation from top quark:}}
The production of a Higgs boson through this channel is relevant for low mass searches, leading to
\begin{equation}
\sigma_{\mathrm{NLO}}(Ht\overline{t})=(1-\alpha_{\partial\phi}v^2) \, \sigma^{\mathrm{SM}}_{\mathrm{NLO}}(Ht\overline{t})\,.
\end{equation}
The LO cross section has been obtained with the help of the program HQQ~\cite{Spiraurl}, further dressed with a K-factor encapsulating the increase of the cross section due to NLO corrections~\cite{Dittmaier:2011ti,Beenakker:2002nc,Reina:2001sf,Dawson:2002tg}.

\end{description}

\begin{figure}[tp]
\begin{center}
\epsfig{figure=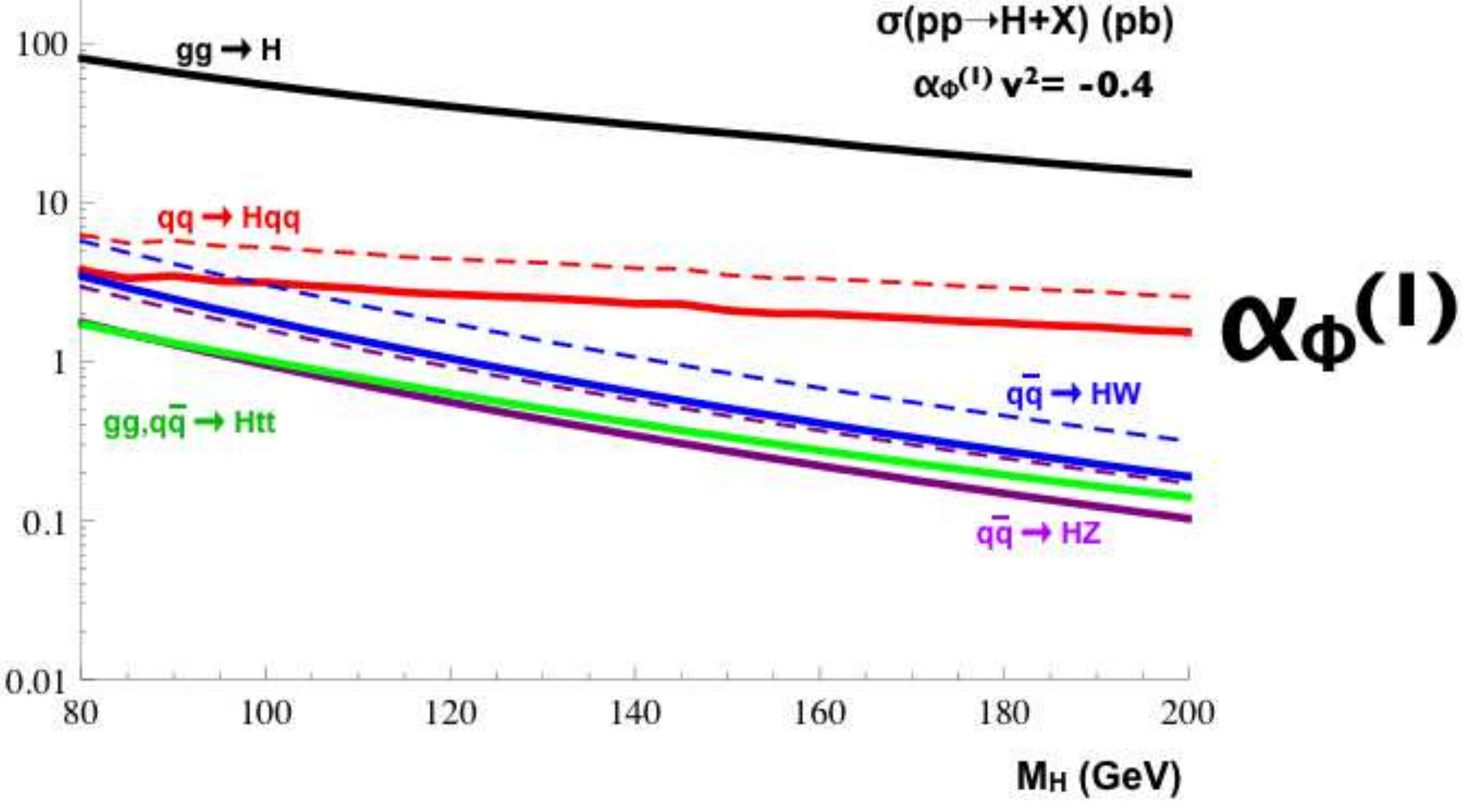,width=8.4cm}
\hspace{-0.30cm}
\epsfig{figure=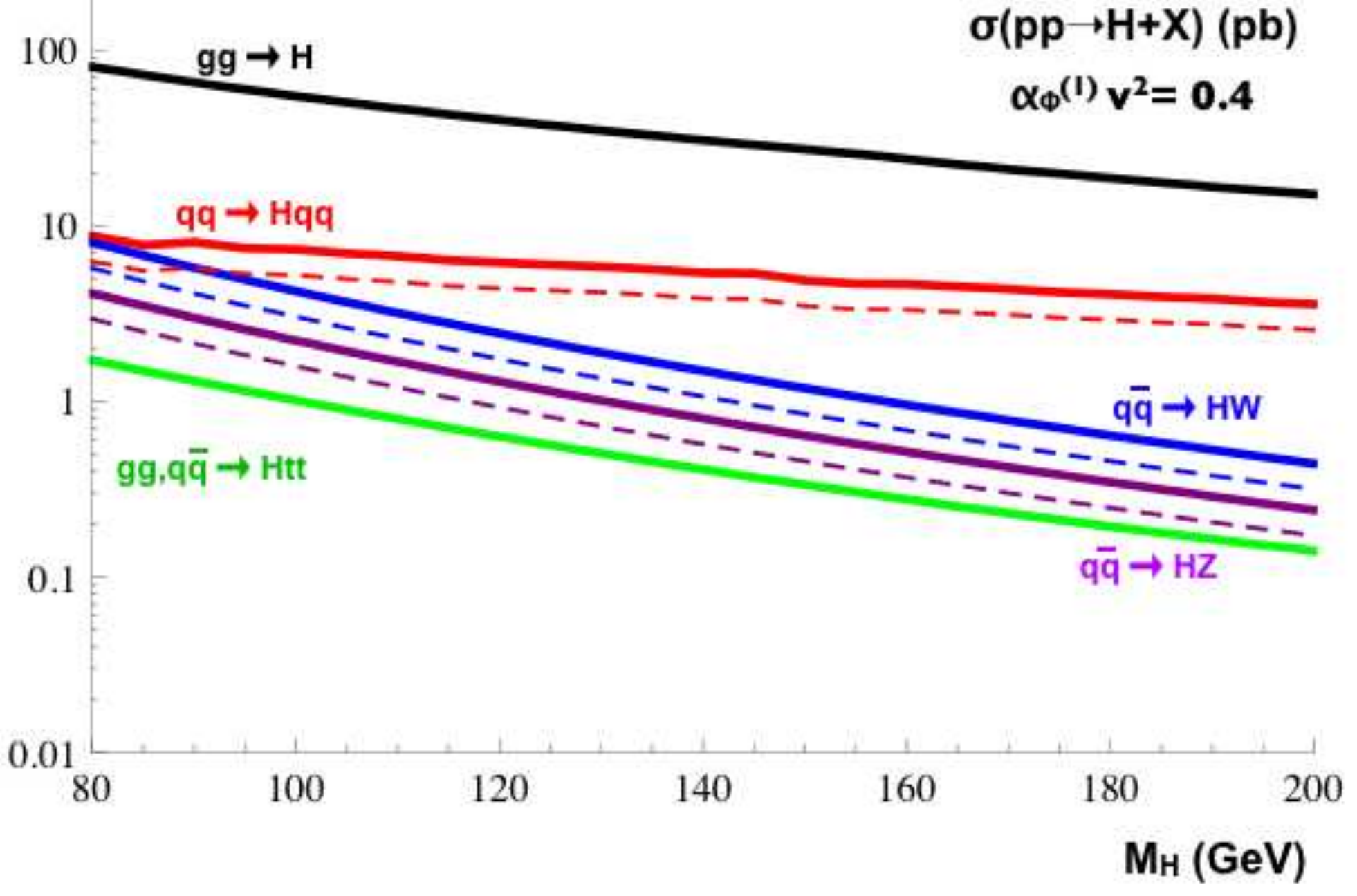,width=7.05cm}
\caption{\label{Prodxf1}
Production cross sections of the Higgs boson at the LHC as a function of the Higgs mass $M_H$ in the presence of $\mathcal{O}_{\phi}^{(1)}$, where $\alpha_{\phi}^{(1)}v^2=-0.4$ (left panel) and $\alpha_{\phi}^{(1)}v^2=0.4$ (right panel). The dashed curves represent the corresponding SM predictions (only visible if there are deviations from the SM).}
\end{center}
\end{figure}

In summary, $\mathcal{O}_{\partial\phi}$ corrects all cross sections by the same factor $1-\alpha_{\partial\phi}v^2$, and thus it leads only to an overall rescaling. On the other hand, for $\mathcal{O}_{\phi}^{(1)}$,   vector boson fusion production and  Higgsstrahlung are the only production mechanisms modified. We show in Fig.~\ref{Prodxf1} the production cross sections of the Higgs boson at the LHC as a function of the Higgs mass $M_H$ in the presence of $\mathcal{O}_{\phi}^{(1)}$, where $\alpha_{\phi}^{(1)}v^2=-0.4$ (left panel) and $\alpha_{\phi}^{(1)}v^2=0.4$ (right panel).
The SM reference curves are shown as dashed curves.
For negative values of  $\alpha_{\phi}^{(1)}$ (left panel), the Higgsstrahlung processes are suppressed, which means that the radiation of a Higgs boson off top quarks becomes as important. For positive values, vector boson fusion and Higgsstrahlung production increase by a factor up to $2.5$. However, gluon fusion remains to be the dominant production channel.


\section{Significance of the Search Channels}
\label{sec:significance}

Here we discuss the significance of the different search channels at the LHC, illustrating it for CMS. First, we show the impact on Higgs discovery searches. Then we discuss the possible discovery of deviations from the SM.

At CMS, for Higgs masses above $125\,\rm{GeV}$, the inclusive production of a Higgs with subsequent decay into 4 leptons via a pair of $Z$ bosons is considered to be the golden discovery channel. When $M_H>2 M_W$, the decay into a pair of $W$ bosons takes over. For low masses, the most promising channel turns out to be the decay of the Higgs boson into a pair of photons. Non-inclusive channels, relying on vector boson fusion are also useful. Higgs production in vector bosons fusion, with decay into $WW$, is a quite efficient channel in the intermediary mass region ($140 - 180 \, \rm{GeV})$. In the low mass region, the channel where the the Higgs decays into a pair of tau leptons can help reaching the $5\sigma$ significance. To summarize we are investigating the following search channels:
\begin{itemize}
\item Inclusive production with decay: \begin{itemize}
\item $H\rightarrow \gamma\gamma$
\item $H\rightarrow ZZ \rightarrow 2e2\mu,\,4e,\,4\mu$
\item $H\rightarrow WW\rightarrow 2\ell2\nu$
\end{itemize}
\item Vector boson production ($qqH$) plus decay:  
\begin{itemize}
\item $H\rightarrow WW \rightarrow \ell\nu jj$
\item $H\rightarrow \tau\tau\rightarrow \ell+ j + E_{\rm T}^{\rm{miss}}$
\end{itemize}
\end{itemize}

\subsection{Impact on Higgs Discovery Searches}

In order to obtain the significance for the most relevant Higgs searches channels at the LHC, we follow the procedure of \Ref~\cite{Espinosa:2010vn} and we refer to the analysis of the CMS collaboration (CMS TDR)~\cite{Ball:2007zza}. For each channel, the number of signal events $s$ and the number of background events $b$ are obtained after the application of experimental cuts: with these number of events,  the significance $S$ is then estimated.
As the effective operators $\mathcal{O}_{\partial\phi}$ and $\mathcal{O}_{\phi}^{(1)}$ modify only the Higgs couplings, the background processes remain as in the SM, \ie, the number of background events $b=b_{\mathrm{SM}}$. The 
number of signal events $s$ in the presence of the effective operators is instead scaled by
$s=\delta\cdot s_{\mathrm{SM}}$ with 
\begin{equation}
\delta=\frac{\sigma(X\rightarrow H)\times \mathrm{BR}(H\rightarrow Y)}{\sigma_{\mathrm{SM}}(X\rightarrow H)\times \mathrm{BR_{\mathrm{SM}}}(H\rightarrow Y)}\, .
\end{equation}
Here $\sigma(X\rightarrow H)$ denotes the production cross section of the Higgs boson via the process $X$ and $ \mathrm{BR}(H\rightarrow Y)$ denotes the decay of the Higgs boson into a given final state $Y$. This means that the product of production and decay enters the different search channels, and therefore also the corresponding modifications resulting from  BSM physics. The values of $s_{\mathrm{SM}}$ and $b_{\mathrm{SM}}$ can be obtained from the CMS analysis~\cite{Ball:2007zza}.

\begin{figure}[t!]
\begin{center}
\epsfig{figure=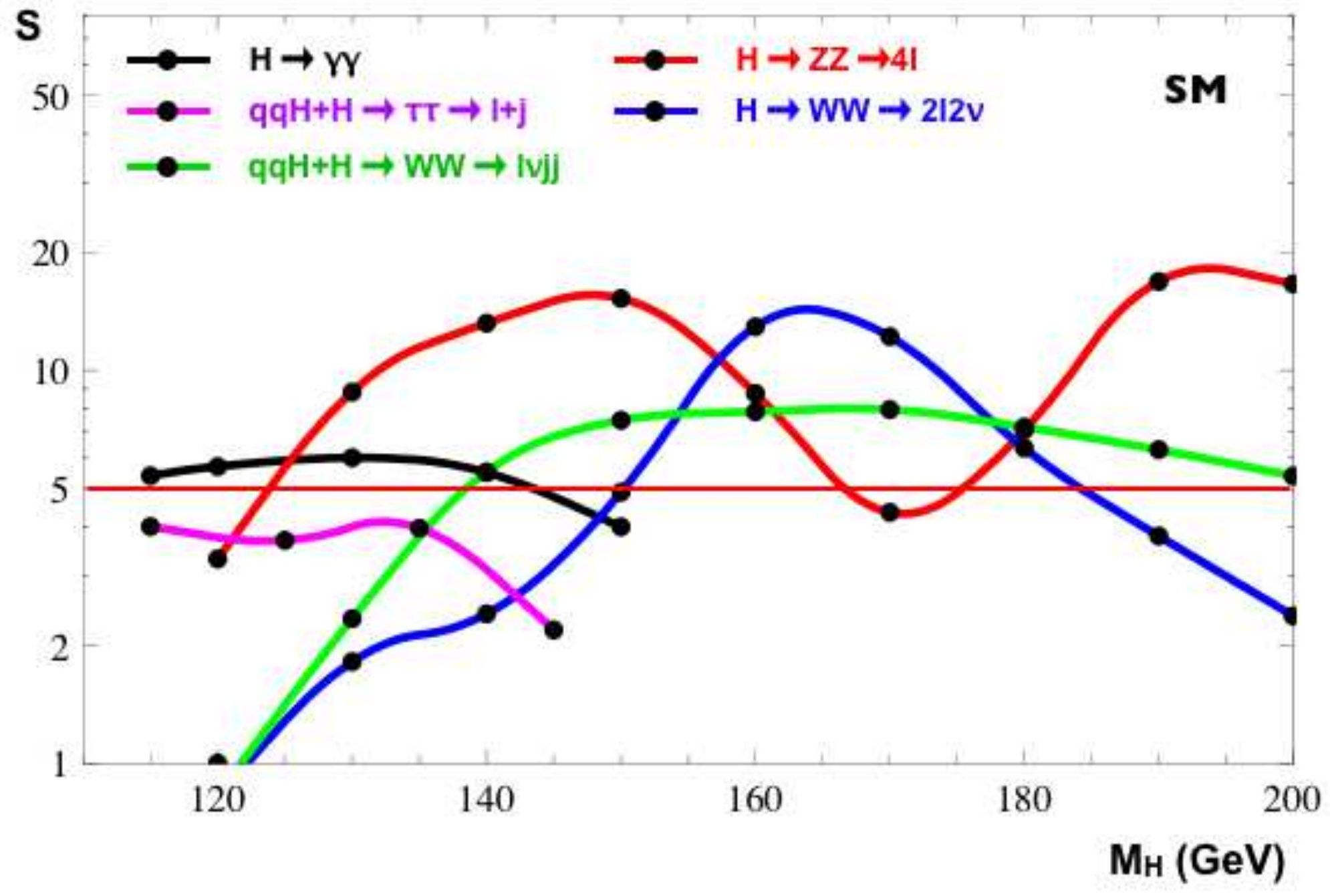,width=7.3cm}\\
\epsfig{figure=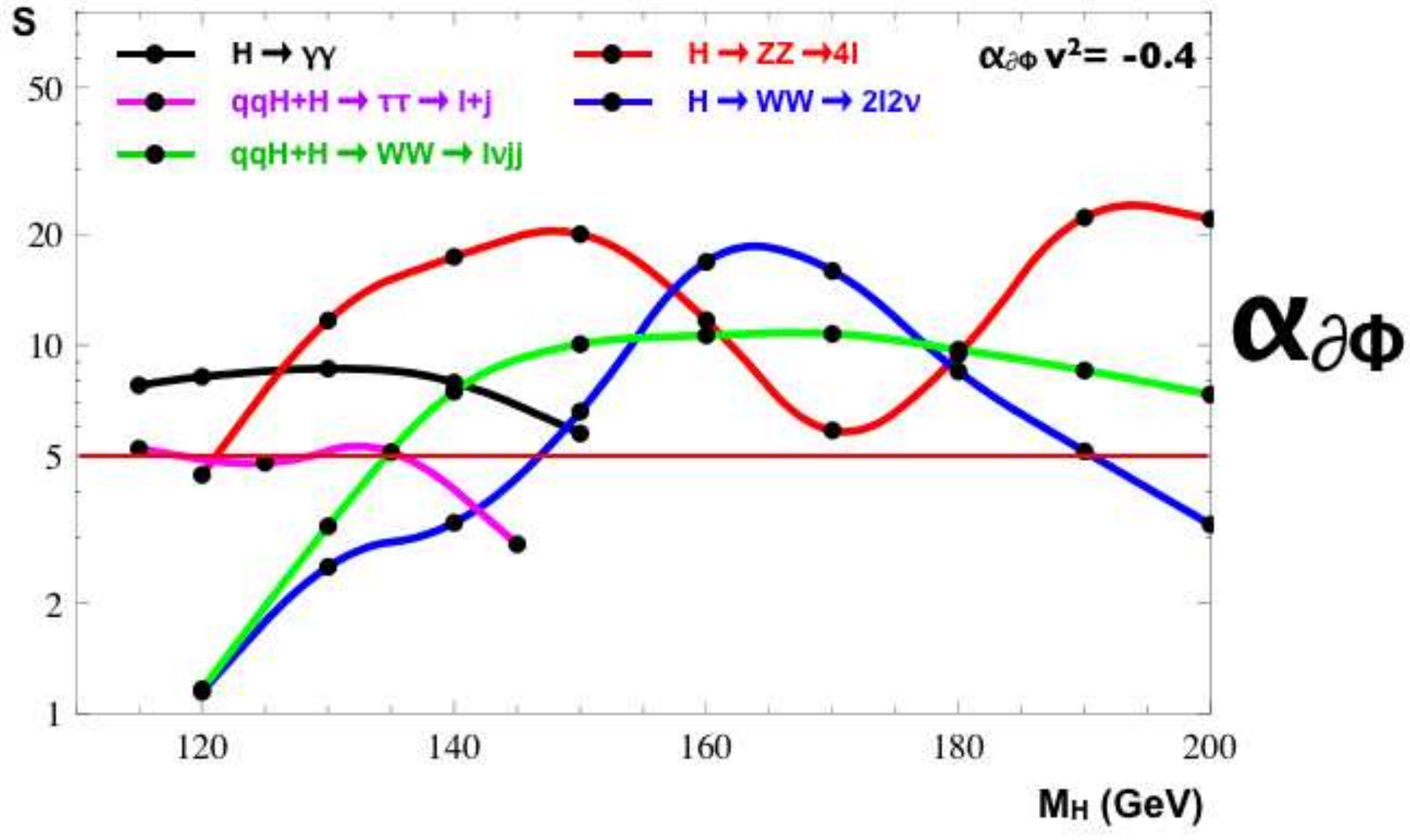,width=8.04cm}
\hspace*{-0.30cm}
\epsfig{figure=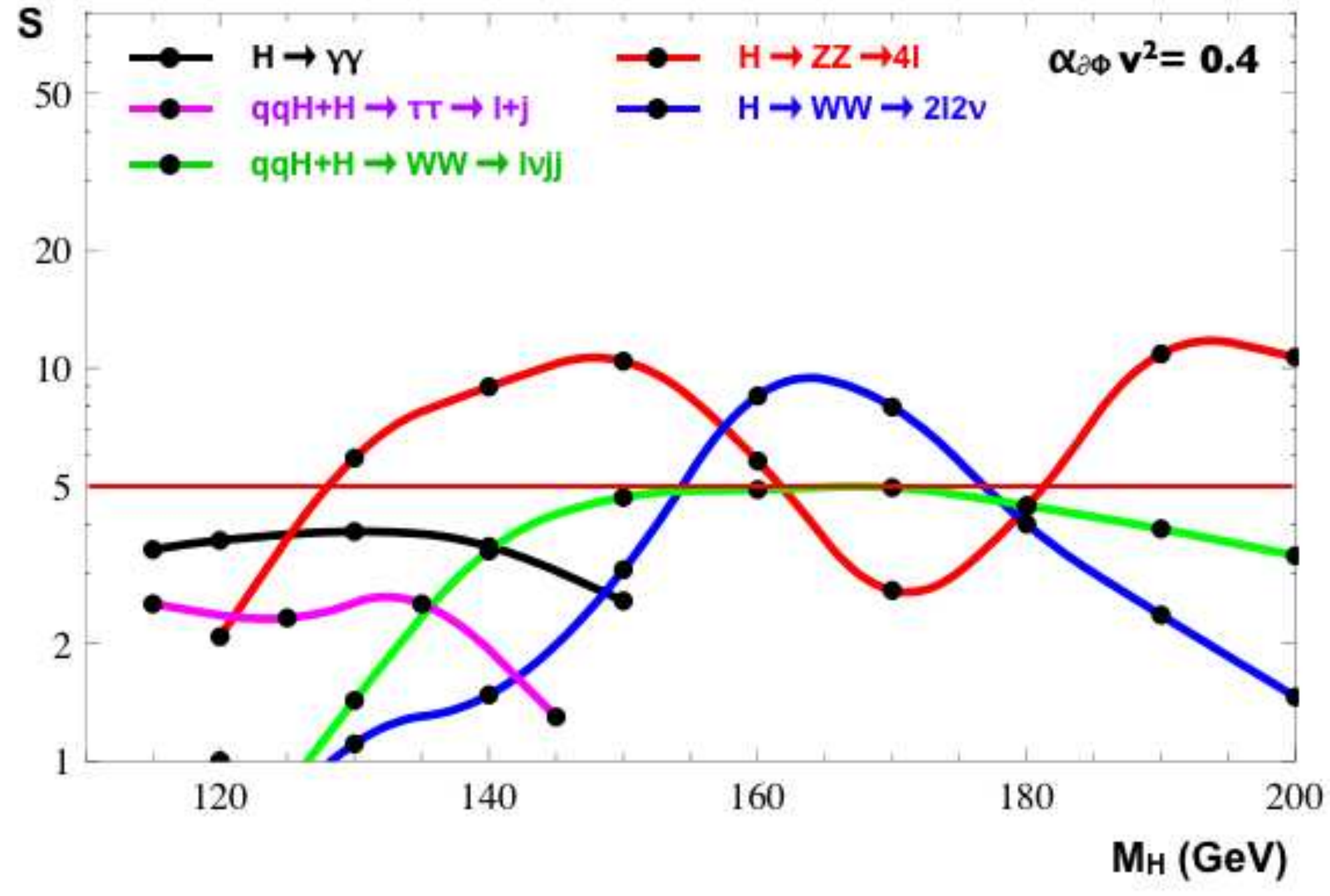,width=7.1cm}\\
\epsfig{figure=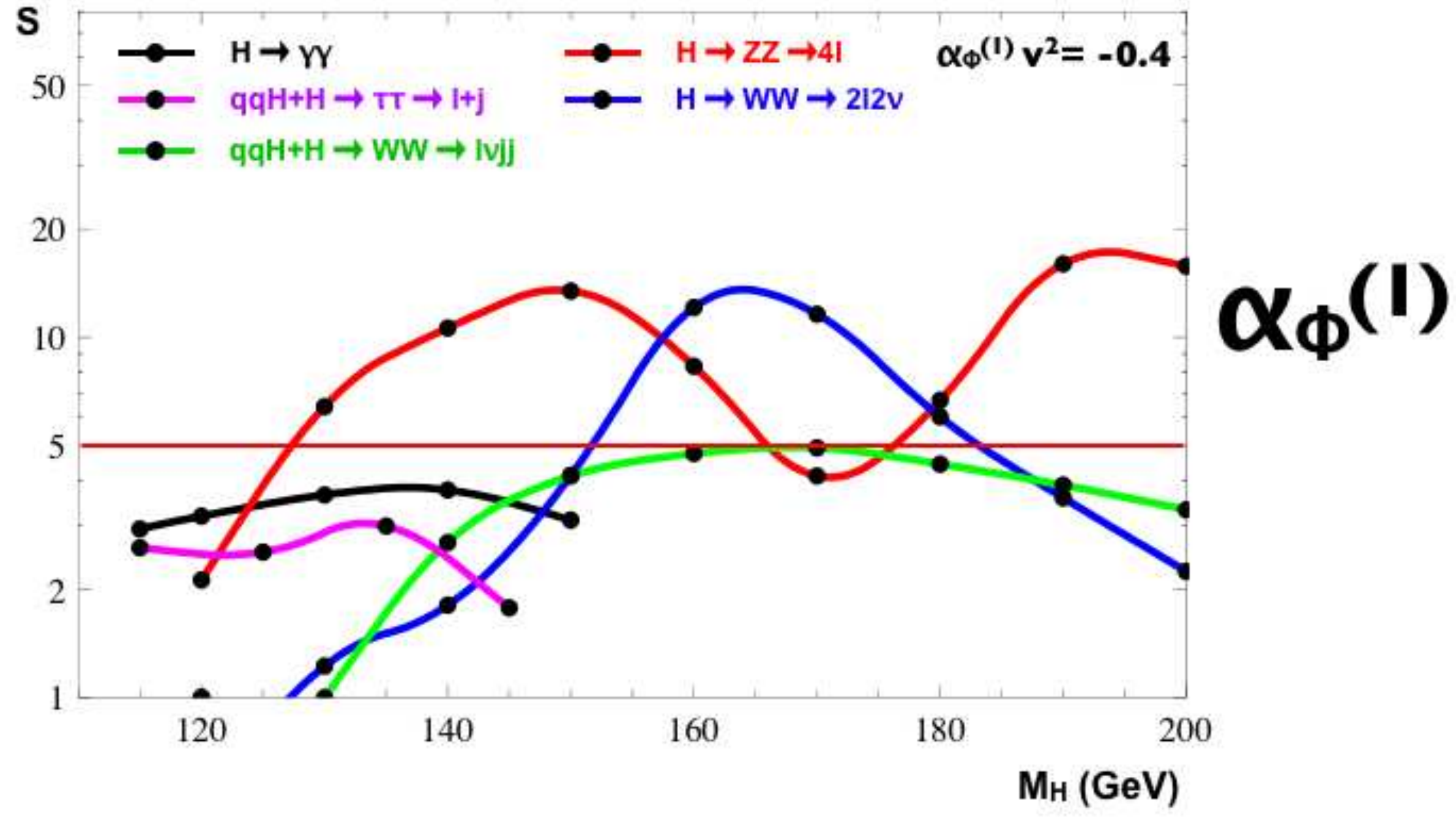,width=8.4cm}
\hspace*{-0.30cm}
\epsfig{figure=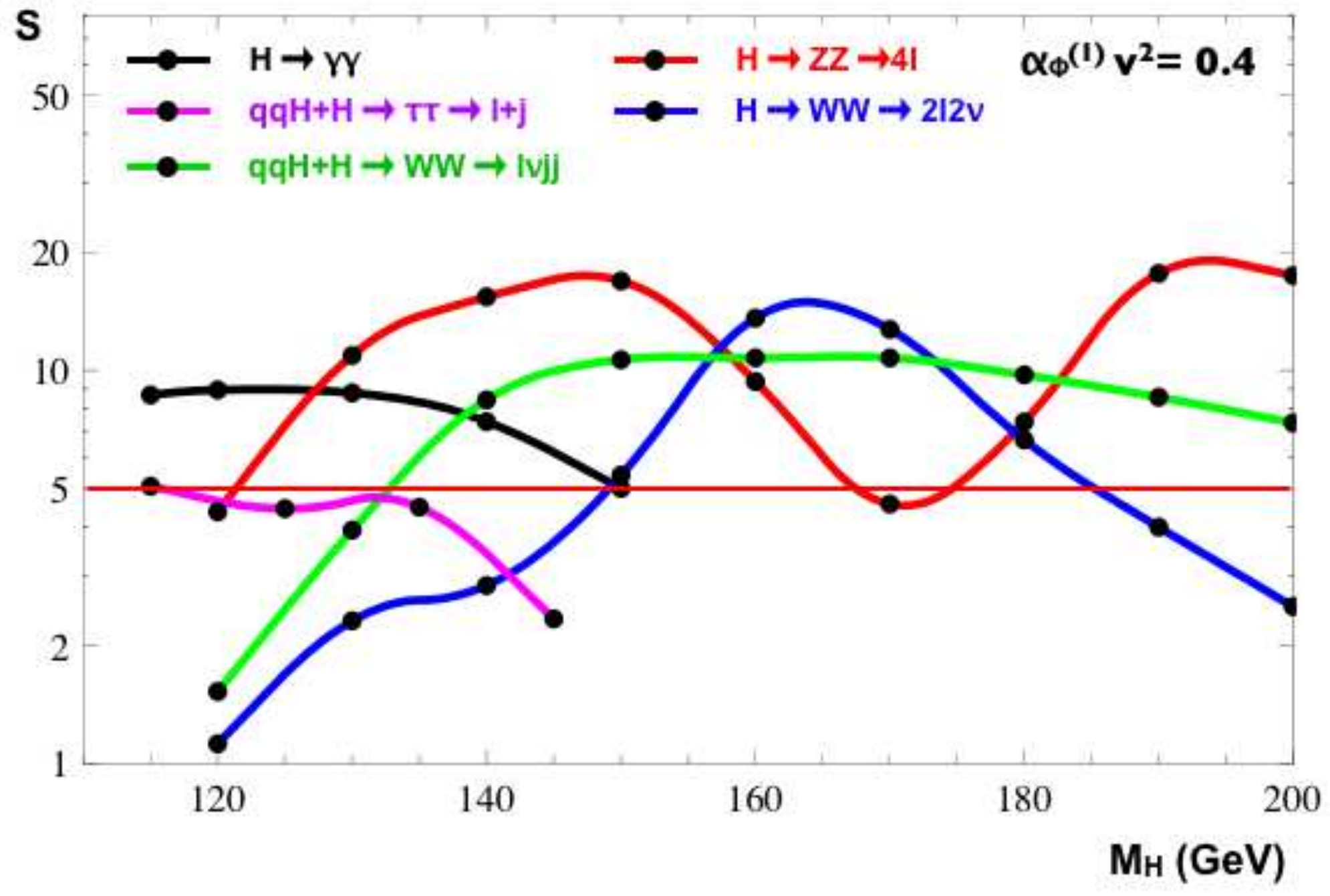,width=7.05cm}
\caption{\label{Sigxdf}
The significances of the different Higgs searches channels at CMS as a function of the Higgs boson mass in the cases: SM (top),  $\alpha_{\partial\phi}v^2=-0.4$ (middle left), $\alpha_{\partial\phi}v^2=0.4$ (middle right), $\alpha_{\phi}^{(1)}v^2=-0.4$ (lower left) and $\alpha_{\phi}^{(1)}v^2=0.4$ (lower right)}
\end{center}
\end{figure}

Following Ref.~\cite{Espinosa:2010vn}, we have used various definitions for the significance, depending on the process analyzed, in order to remain as close as possible to the CMS results for the case of the SM. 
The conventions in Ref.~\cite{Espinosa:2010vn} have been used. We have set the integrated luminosity to $\int \mathcal{L}=30\, \mathrm{fb}^{-1}$, to facilitate  the  comparison with previous studies.
The significances of the SM Higgs boson searches at  a $14\,\mathrm{TeV}$ LHC are depicted in Fig.~\ref{Sigxdf} (upper panel), which also illustrates that different search channels for a Higgs bosons at the LHC cover different mass ranges.
As explained above, $\mathcal{O}_{\partial\phi}$ modifies all Higgs couplings by the same factor $1-\alpha_{\partial\phi}v^2$. In consequence, all significances get enhanced (depleted) with respect to the SM ones for negative (positive) values of   
$\alpha_{\partial\phi}$. 
For $\mathcal{O}_{\phi}^{(1)}$, instead, 
given the positive sign of the $\alpha_{\phi}^{(1)}$ contribution to the couplings, see Secs.~\ref{section:decay} and~\ref{sec:production}, the general trend expected is an increase (decrease) with respect to the SM predictions for positive (negative) values of $\alpha_{\phi}^{(1)}$.

The analysis of the significances as a function of the Higgs mass,  for
each of the different search channels separately,  can be found in App.~\ref{sec:significances}. Fig.~\ref{Sigxdf}  summarizes the results for $\alpha_{\partial\phi}$ and $\alpha_{\phi}^{(1)}$, respectively. 
When the new physics is induced exclusively by the operator $\mathcal{O}_{\partial\phi}$, see middle row, 
for  $\alpha_{\partial\phi}<0$ the enhancement with respect to the SM is especially strong for $M_H>150\,\mathrm{GeV}$. Nevertheless, for Higgs boson masses between $160-180\,\mathrm{GeV}$, in general only positive values of $\alpha_{\partial\phi}$ are allowed  because of the  Tevatron bounds (\cf, Fig~\ref{LEPTEV}). For the chosen value $\alpha_{\partial\phi} v^2= 0.4$, the depletion of all production processes with respect to the SM predictions is such that reaching a $5\sigma$ significance becomes difficult in the low mass region in the early stages of LHC, in which the $H\rightarrow\gamma\gamma$ channel gets deteriorated.  Notice that the analysis for this operator is equivalent to the one performed by Espinosa et al.~\cite{Espinosa:2010vn}, for the composite Higgs model named MCHM4, and we checked that our results are consistent with theirs.

If, on the other hand, only the operator $\mathcal{O}_{\phi}^{(1)}$ is
non-vanishing -- see lower row of Fig.~\ref{Sigxdf} -- the qualitative
behavior is inverted.  For negative values of $\alpha_{\phi}^{(1)}$, a
significant increase of statistics (or the combination of different
channels) is needed for  a $5\sigma$ significance in the low mass
region, for the chosen parameter value. In the intermediate mass range,
below the gauge threshold, a soft diminution of the significance of the
different search channels is observed, while above the gauge threshold,
where the branching ratio $H\rightarrow WW$ is almost equal to $1$, the
significances almost equal those of the SM. This does not apply to the
$qqH + H\rightarrow WW \rightarrow \ell\nu jj$ channel,  which is not
inclusive, \ie, relies only on  vector boson fusion production of the
Higgs boson. For this channel, the significance in the high mass region
is significantly lower than in the SM for negative values of
$\alpha_{\phi}^{(1)}$. For positive values of 
$\alpha_{\phi}^{(1)}$, the enhancement induced is such that 
even the $H\rightarrow \tau\tau$  channel might reach 
$5\sigma$ significance. 
The $qqH + H\rightarrow WW \rightarrow \ell\nu jj$ channel gets also 
substantially enhanced  and can even compete with the 
$H\rightarrow WW \rightarrow 2\ell2\nu$ channel which, 
together with the $H\rightarrow ZZ\rightarrow4\ell$ channel, 
is only slightly enhanced in comparison.

To summarize, the early discovery of the Higgs boson at LHC with
moderate luminosities 
is a relatively robust prediction even in the presence of physics BSM, 
unless $M_H \lesssim 130 \, \mathrm{GeV}$.

Let us consider now the case in which the BSM physics may induce
simultaneously both 
$\mathcal{O}_{\partial\phi}$ and $\mathcal{O}_{\phi}^{(1)}$ interactions.
Eqs.~(\ref{HZZm}) to~(\ref{HWWm}) illustrate that only the couplings of the Higgs boson to massive gauge bosons are sensitive to both operators, via the  factor $(1+\alpha_{\phi}^{(1)}v^2-\alpha_{\partial\phi}v^2)$.  This obviously implies that significant departures from the results obtained above  may only happen when:
{\begin{description}
\item[ $\alpha_{\partial\phi}=-\alpha_{\phi}^{(1)}$:] the Higgs coupling to massive gauge bosons will be as predicted in the SM case, while the coupling to gluons and fermions will be modified by the non-zero value of $\alpha_{\partial\phi}$. 
\item [$(1+\alpha_{\phi}^{(1)}v^2-\alpha_{\partial\phi}v^2)=0$:] as we remain in the perturbative regime, such a cancellation occurs only  at the point $(\alpha_{\partial\phi}v^2,\alpha_{\phi}^{(1)}v^
 2)=(0.5,-0.5)$, 
at the limit of the perturbative region. For such an
extreme case, the Higgs is no longer coupled to the $W$ or
$Z$ bosons and Higgs searches are very compromised.
In the corner of the parameter space close to that point, only low mass searches get really affected.
 In both cases, comparison between inclusive and non-inclusive channels should allow to detect if some interplay between both effective operators is at work.
\end{description}}

\subsection{Discovery of Deviations from the SM at the LHC}
\label{sec:discdev}

\begin{figure}[t]
\begin{center}
\epsfig{figure=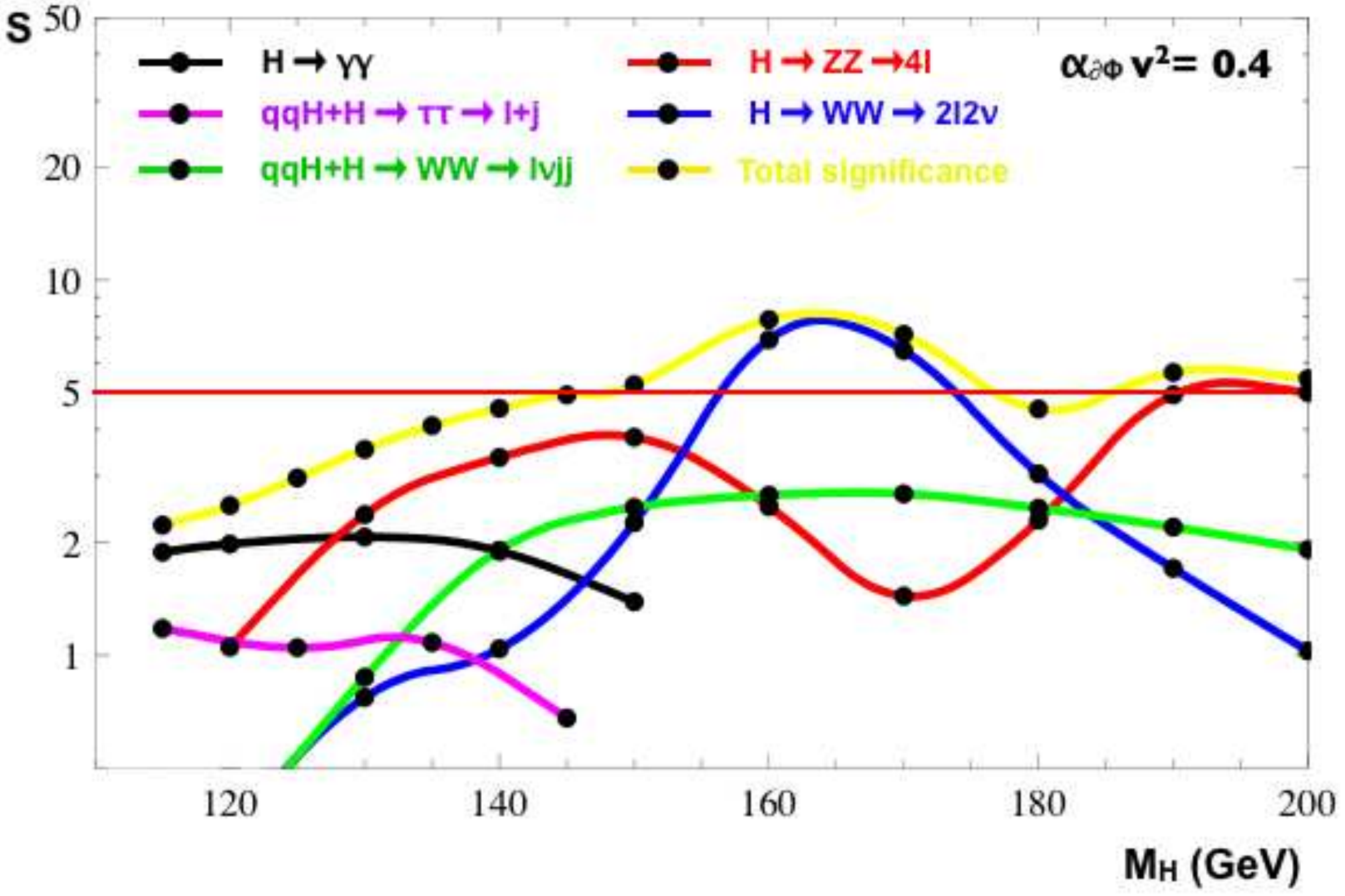,width=7.3cm}
\hspace*{0.5cm}
\epsfig{figure=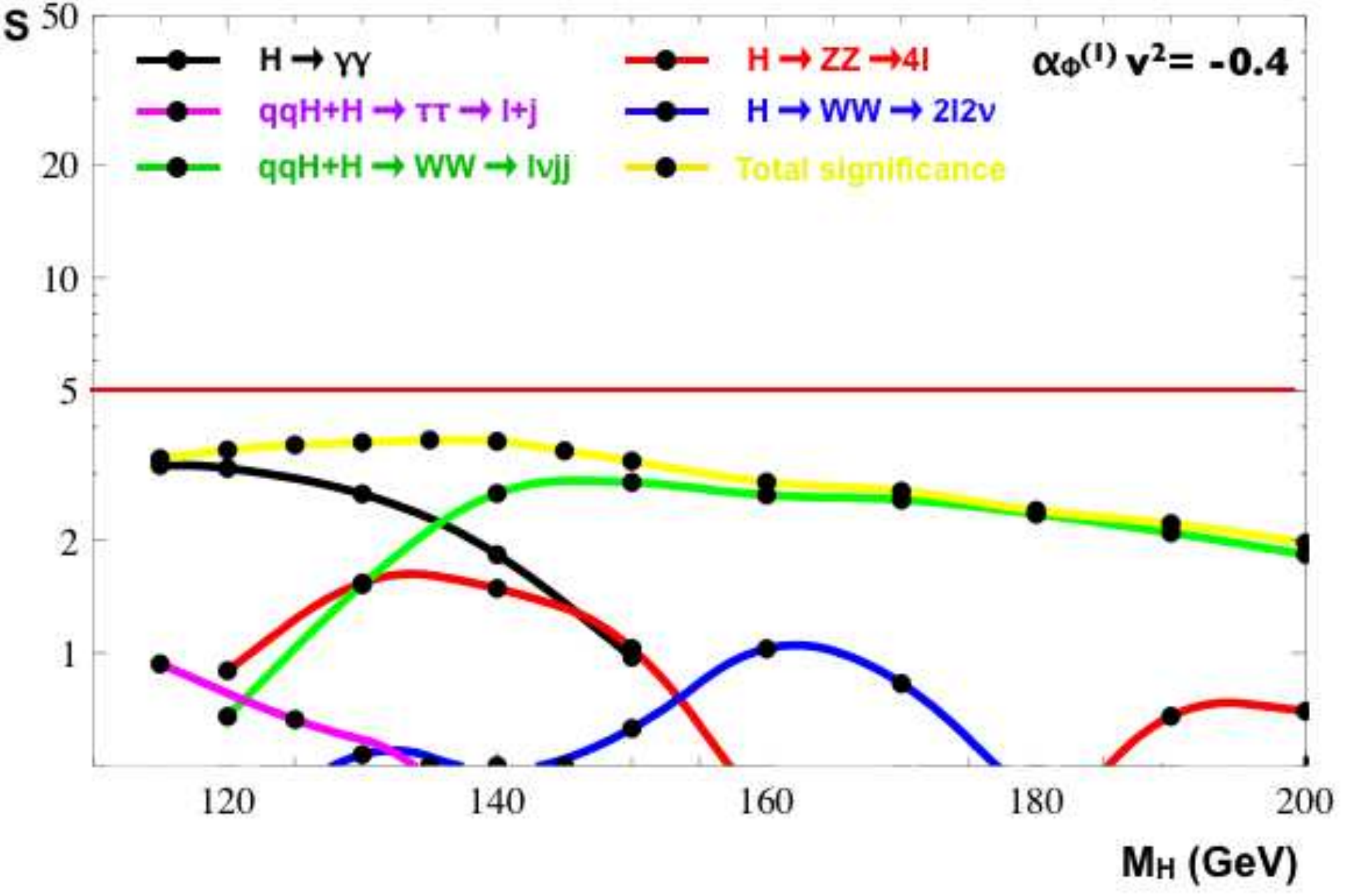,width=7.3cm}
\caption{\label{sens}
Significance for the discovery of the effective coefficient $\alpha_{\partial\phi}$ (left) and $\alpha_{\phi}^{(1)}$ (right) as a function of the Higgs boson mass from the different searches channels. Here the a luminosity of $30~\mathrm{fb}^{-1}$ is used, and a simulated value $\alpha_{\partial\phi} v^2=0.4$ (left panel) and $\alpha_{\phi}^{(1)} v^2=-0.4$ (right panel).  }
\end{center}
\end{figure}

We have just seen how the presence of effective operators could modify
the searches for a Higgs boson at the LHC. It is also interesting to
know to what extent the LHC will be able to discover the effective
coefficients $\alpha_{\partial\phi}$ and $\alpha_{\phi}^{(1)}$. To do
that, we have studied the significance of the rejection of the SM theory
over the data one would get in the presence of the effective
operators. We used the same formulae used in the previous section for
the signal rates of the different channels. 
Fig.~\ref{sens} shows the result for the cases $\alpha_{\partial\phi} v^2=0.4$ (left) and $\alpha_{\phi}^{(1)} v^2=-0.4$ (right).
Note that the sensitivity for a negative coefficient is almost the same as the one for positive coefficient.

First of all, note that an early high significance discovery is only possible for $\alpha_{\partial \phi}$ if $M_H \gtrsim 160 \, \mathrm{GeV}$ (and $|\alpha_{\partial \phi}|$ large enough), whereas it is not possible for $\alpha_\phi^{(1)}$. However, for about $200 \, \mathrm{fb^{-1}}$ (two years at design luminosity), it can be roughly estimated that either of these two effects can be discovered for $|\alpha|v^2 \gtrsim 0.4$, barring cancellations, for any allowed value of $M_H$. In the extreme limit, such as after 10 years running at the design luminosity ($1000 \, \mathrm{fb^{-1}}$), we expect that values $|\alpha| v^2\gtrsim 0.17$ may be discovered for any allowed value of $M_H$. In the most optimistic case ($M_H \simeq 170 \, \mathrm{GeV}$), we expect a discovery for $|\alpha_{\partial \phi}| v^2 \simeq 0.04$ in that case.  These conclusions are only true if only one of the two effects is considered independently. If, however, both effects are present at the same time, 
 there may be
  cancellations or additions of the effects, see earlier formulas. Note that these results are coherent with previous studies such as that in Ref.~\cite{Lafaye:2009vr}, where it has been shown that the expected sensitivity in deviations of the Higgs SM couplings is around $20\%$.

 If a deviation from the SM is discovered and needs to be interpreted, one needs to know from which operator it comes from. The discrimination power between  $\alpha_{\partial \phi}$ and $\alpha_\phi^{(1)}$ relies on the relative contribution to different search channels: while $\alpha_{\partial \phi}$ affects all channels in the same way, see Fig.~\ref{Sigxdf} (middle row), $\alpha_\phi^{(1)}$ leads to relative shifts among the different contributions, see Fig.~\ref{Sigxdf} (lower row).
 In particular, non-inclusive search channels that rely only on vector boson fusion would be very useful. For intermediate and high Higgs mass, the event rate in the  $qqH+H\rightarrow WW \rightarrow \ell\nu jj$ channel  compared to the  $H\rightarrow ZZ \rightarrow 4\ell$ and $H\rightarrow WW \rightarrow 2\ell 2\nu$ channels should provide a test for $\alpha_{\phi}^{(1)}$ versus $\alpha_{\partial\phi}$. In the low mass range, the situation might be more difficult, since the inclusive $H\rightarrow\gamma\gamma$ channel is sensitive to both coefficients. However the $qqH+ H\rightarrow\tau^+\tau^-\rightarrow \ell+j+E_{\mathrm{T}}^{\mathrm{miss}}$ channel should help to disentangle the two effective operators considered. Such a discrimination may be possible with higher statistics at later stages of the LHC operation.  
In addition, it may be interesting to know the  sign of the deviation from the SM, as this may be indicative for certain models (see next section).
This  is essentially always possible once a departure from the SM prediction is detected, because the deviations are sensitive to the sign of $\alpha$ and not only $|\alpha|$. The discovery reach for a particular sign is comparable to that of the discovery reach for $|\alpha|$. Measuring the sign in the presence of both effects relies, again, one the relative contribution from  different channels, as discussed above.

\section{Interpretation of Deviations from the SM at Tree Level}
\label{sec:theory}

 While the earlier sections of this study are independent of the high energy theories leading to the effective operators, we here interpret possible deviations from the Standard Model if these effective operators are generated at tree level via the exchange of heavy new mediators. Since the test of $\mathcal{O}_\phi$  requires beyond-LHC technology and $\alpha_\phi^{(3)}$ is strongly constrained by EWPT, only $\mathcal{O}_{\partial \phi}$ and $\mathcal{O}_{\phi}^{(1)}$ can be probed at the LHC, as argued above. However, we will consider also $\mathcal{O}_{\phi}^{(3)}$ in this section, since its impact on EWPT  will immediately indicate which  mediators cannot be expected to have an observable LHC impact via   $\mathcal{O}_{\partial \phi}$ and $\mathcal{O}_{\phi}^{(1)}$.

For the analysis, we first of all find all possible tree-level decompositions of the operators in Eqs.~(\ref{ophidphi}) and~(\ref{ophi1ophi3}), from which we obtain a list of possible mediators; see Refs.~\cite{Gavela:2008ra,Bonnet:2009ej} for the applied techniques. As the next step, we formulate the fundamental Lagrangian with all of these mediators and all relevant interactions. Finally, we re-integrate out these mediators  simultaneously,  to verify  that all multi-mediator interactions have been taken into account. For the scalar mediators, this procedure is quite straightforward. For the vector mediators, however, the result depends on whether the interactions are introduced such that 1) the vectors are gauge fields~\cite{Arzt:1994gp}
 or 2) the vectors are not gauge fields (which implies that the
 Lagrangian results from a broken gauge symmetry and is not
 fundamental). In the first case, only a singlet or a triplet
 (hypercharge neutral)
 vector boson are allowed as mediators, whereas in the second case, other
 vector mediators are possible.
 We will focus on  singlet and triplet
 vectors, and show the differences between gauge and non-gauge
 interactions where applicable.
We find the following list of mediators:
\begin{gather}
S({\bf 1}^{s}_{0}),
\quad
\varphi ({\bf 2}^{s}_{1/2}),
\quad
\Delta^{a} ({\bf 3}^{s}_{0}),
\quad
\Delta_{1}^{a} ({\bf 3}^{s}_{1}),
\quad
V_{\rho} ({\bf 1}^{v}_{0}),
\quad
U^{a}_{\rho} ({\bf 3}^{v}_{0}) \, ,
\label{equ:allmed}
\end{gather}
where we have assigned symbols to the mediators and list the SM quantum numbers in brackets in the form ${\bf X}^{\mathcal{L}}_Y$, where
\begin{itemize}
\item
 ${\bf X}$ denotes the SU(2) nature, i.e., singlet ${\bf 1}$, doublet ${\bf 2}$, or triplet ${\bf 3}$.
\item
     $\mathcal{L}$ refers to the Lorentz nature, i.e., scalar ($s$) 
     and vector ($v$).
\item
 $Y$ refers to the hypercharge $Y=Q-I^W_3$.
\end{itemize}
Note that we find decompositions with up to three  mediators which
differ by their Lorentz and/or SM quantum numbers. In addition, we
introduce the doublet scalar $\varphi$ without kinetic and mass mixing
with the SM Higgs doublet.

The primary goal of this section is to clarify  
the minimum set of renormalizable interactions necessary to generate the effective operators under study in this paper.
This allows  to address the following question: 
what can be predicted for physics at high energy scales
from the effective interactions, if they are discovered?
Nevertheless, a word of caution is convenient. The fields in
\equ{allmed} may come from numerous models at high energies
(see \eg,
Refs.~\cite{Georgi:1985nv,Chanowitz:1985ug,Gunion:1989ci,Gunion:1990dt,Logan:2010en,Pati:1974yy,Mohapatra:1974hk,Senjanovic:1975rk}
for the triplet scalars)
and the list of couplings discussed here may not 
cover the full set of interactions of a given concrete high-energy model.


\subsubsection*{One mediator cases}

Let us first discuss the case when only one mediator is present.
Assuming that the vectors are gauge fields, the relevant part of the
fundamental Lagrangian\footnote{
Note that these minimal set-ups might lead to tree-level unitarity
violation in the different scattering amplitudes.  However, unitarity
consideration should be worked out in complete models where the fields
under study are embedded. Such study is beyond the scope of this paper,
but some example can be found for $Z'$ model \cite{Basso:2011na} or
Higgs triplet model~\cite{Aoki:2007ah} for example.} 
leading to $\alpha_{\partial \phi}$, $\alpha_\phi^{(1)}$, and $\alpha_\phi^{(3)}$ is
\begin{eqnarray}
\mathcal{L}_{\mathrm{LHC}}
& =&
\frac{1}{2} 
(\partial_{\rho} S)
(\partial^{\rho} S)
-
\frac{1}{2}
m_{S}^{2}
S^{2}
+
\mu_{S}
(\phi^{\dagger} \phi) S
  \label{equ:laglhc}
 \\
& + & \frac{1}{2}
(D_{\rho} \Delta)^{a}
(D^{\rho} \Delta)^{a}
-
\frac{1}{2} 
m_{\Delta}^{2}
\Delta^{a} \Delta^{a}
+
\mu_{\Delta}
(\phi^{\dagger} \tau^{a} \phi)
\Delta^{a} 
\nonumber
\\
& + &
(D_{\rho} \Delta_{1})^{\dagger a}
(D^{\rho} \Delta_{1})^{a}
-
m_{\Delta_{1}}^{2}
\Delta_{1}^{\dagger a}
\Delta_{1}^{a}
 + 
\left[
\mu_{\Delta_{1}}
(\phi {i} \tau^{2} \tau^{a} \phi)
\Delta_{1}^{\dagger a}
+
{\rm h.c.}
\right]
\nonumber 
\\
& - &
\frac{1}{4} 
V_{\rho \sigma} V^{\rho \sigma}
+
\frac{1}{2} m_{V}^{2} V_{\rho} V^{\rho}
-
{i} g_{V}
V_{\rho}
[(D^{\rho} \phi)^{\dagger} \phi - \phi^{\dagger} (D^{\rho} \phi)] 
 \nonumber
\\
& - &
\frac{1}{2} 
U^{a}_{\rho \sigma} U^{a \rho \sigma}
+
\frac{1}{2}
m_{U}^{2} U^{a}_{\rho} U^{a \rho}
-
{i} 
\frac{g_{U}}{2}
U^{a}_{\rho}
\left[
(D^{\rho} \phi)^{\dagger} 
{\tau^{a}} \phi
-
\phi^{\dagger }
{\tau^{a}}
(D^{\rho} \phi)
\right] \, ,
\nonumber
\end{eqnarray}
where $\tau^{a}$ $(a=1,2,3)$ are Pauli matrices for 
SU(2) gauge symmetry,
 $V_{\rho \sigma}$ and $U_{\rho \sigma}^{a}$ 
denote the field strength tensors for 
the  SU(2)-singlet $V_{\rho}$ and SU(2)-triplet $U_{\rho}^{a}$ vector fields, $ \Delta$ and $\Delta_1$ denote scalar SU(2)-triplets with different hypercharge, and $S$ stands for scalar singlet.  The scalar doublet $\varphi$ does not appear here since  it does not contribute
to the LHC-observable operators, $\mathcal{O}_{\partial \phi}$ and $\mathcal{O}_{\phi}^{(1)}$ but only to $\mathcal{O}_\phi$ (see \App~\ref{App:fulldecom}). 

After integrating out the mediators, we obtain
\begin{eqnarray}
\mathscr{L}_{\text{LHC}}^{\mathrm{eff}}
& \supset &
\left[
\frac{\mu_{S}^{2}}{m_{S}^{2}}
+
\frac{\mu_{\Delta}^{2}}{m_{\Delta}^{2}}
+
4 
\frac{|\mu_{\Delta_{1}}|^{2}}{m_{\Delta_{1}}^{2}}
\right]
\frac{1}{2}
(\phi^{\dagger} \phi)^{2}
\nonumber 
\\
&
+
&
\left[
\frac{\mu_{S}^{2}}{m_{S}^{4}}
+
\frac{\mu_{\Delta}^{2}}{m_{\Delta}^{4}}
+
\frac{g_{V}^{2}}{m_{V}^{2}}
+
\frac{g_{U}^{2}}{4m_{U}^{2}}
-
\frac{\mu_{S}^{2} \mu_{VS}^{2}}{m_{S}^{4} m_{V}^{2}}
\right]
\mathcal{O}_{\partial \phi}
 \nonumber \\
& + &
2
\left[
\frac{\mu_{\Delta}^{2}}{m_{\Delta}^{4}}
+
2
\frac{|\mu_{\Delta_{1}}|^{2}}{m_{\Delta_{1}}^{4}}
-
\frac{g_{U}^{2}}{4 m_{U}^{2}} 
\right]
\mathcal{O}_{\phi}^{(1)} 
\nonumber \\
& -& 
2
\left[
\frac{\mu_{\Delta}^{2}}{m_{\Delta}^{4}}
-
2
\frac{|\mu_{\Delta_{1}}|^{2}}{m_{\Delta_{1}}^{4}}
+
\frac{g_{V}^{2}}{m_{V}^{2}}
\right]
\mathcal{O}_{\phi}^{(3)} \, .
\label{equ:effresult}
\end{eqnarray}
Integrating out the heavy fields we not only obtain $d=6$ operators but also a $d=4$ one : $\mathcal{O}_{\phi}^{d=4}=(\phi^{\dagger}\phi)^2$. The presence of such operator can potentially affect  the results shown in Sec.~\ref{sec:SMLag}. The vev receives an extra contribution from this operator. Eqs.~(\ref{vev}) is modified as
\begin{equation}
v^2=v_0^2(1+\alpha_{\phi}\frac{v_0^2}{4\lambda_0}-\frac{\alpha_{\phi}^{d=4}}{\lambda_0})\,,
\end{equation}
while the mass of the Higgs is not modified by such operator.
After performing the renormalization through the $Z$-scheme, the contributions of $\mathcal{O}_{\phi}^{d=4}$ to the Higgs couplings disappear, being absorbed by the input parameters. Thus the results shown in the previous sections are still valid in the presence of such a $d=4$ operator.

The operator $\mathcal{O}_\phi$ is also induced by the integration, but it includes more complicated combinations of  interactions and also includes the contribution from additional mediators, as it will be discussed in Sec.~\ref{sec:beyond}. That is why it is not listed in Eq.~(\ref{equ:effresult}). 
From Eq.~\eqref{equ:effresult}, one can now easily read off $\alpha_{\partial
\phi}$, $\alpha_{\phi}^{(1)}$, and $\alpha_{\phi}^{(3)}$ for the single
mediator cases. We list in \Tab~\ref{tab:combin} the coefficients for the individual mediators.
The results for the vector mediators are
consistent with Ref.~\cite{delAguila:2010mx}.
For
the non-gauge vectors, denoted by ${\bf \tilde 1}^v_0$ 
and ${\bf \tilde 3}^v_0$, we use interactions of the form 
\begin{eqnarray}
\mathcal{L}^{\text{non-gauge}}_{V}
&=&
\lambda_{V} V_{\rho} \partial^{\rho} (\phi^{\dagger} \phi)
,\\
\mathcal{L}^{\text{non-gauge}}_{U}
&=&
\lambda_{U} U_{\rho}^{a} [D^{\rho} (\phi^{\dagger} \tau \phi)]^{a},
\end{eqnarray}
instead of the gauge-inspired interactions
shown in Eq.~\eqref{equ:laglhc}.

\begin{table}[t!]
\centering
\begin{tabular}{|l|p{2.8cm}||c|c|c||c|c||c|c|}
\hline
 & & & & & \multicolumn{2}{c||}{Gauge} & \multicolumn{2}{c|}{Non-gauge} \\
Coeff. & Participating in & ${\bf 1}^s_0$ & ${\bf 3}^s_0$ & ${\bf 3}^s_1$ & ${\bf 1}^v_0$  & ${\bf 3}^v_0$  & ${\bf \tilde 1}^v_0$  & ${\bf \tilde 3}^v_0$\\
\hline
$\alpha_{\phi}^{(1)} $ & $HWW$, $HZZ$ & 0 & $2\frac{\mu_{\Delta}^2}{m_{\Delta}^4}$ & $4\frac{|\mu_{\Delta_1}|^{2}}{M_{\Delta_1}^4}$ & 
0 & $- \frac{g_U^2}{2 m_U^2}$ & 0 & $-2 \frac{\lambda_{U}^2}{m_U^2}$ \\
\hline
$\alpha_{\phi}^{(3)}$ & EWPT! & 0 & $-2\frac{\mu_{\Delta}^2}{m_{\Delta}^4}$ & $4\frac{|\mu_{\Delta_1}|^{2}}{m_{\Delta_1}^4}$ & $-2 \frac{g_V^2}{m_V^2}$ & 0 & 0 & $2 \frac{\lambda_{U}^2}{m_U^2}$ \\
\hline
$\alpha_{\partial\phi} $ & $HWW$, $HZZ$, $H \bar f f$ & $\frac{\mu_S^2}{m_S^4}$ & $\frac{\mu_{\Delta}^2}{ m_{\Delta}^4}$ & $0$ & $\frac{g_V^2}{m_V^2}$ & $\frac{g_U^2}{4 m_U^2}$ & $- \frac{\lambda_{V}^2}{m_V^2}$ & $- \frac{\lambda_{U}^2}{m_U^2}$\\
\hline
\end{tabular}
\caption{\label{tab:combin} Individual coefficients 
for the different mediators identified in this section (only single mediator cases). The mediator ${\bf 2}^s_{1/2}$ will not give any contribution here. 
}
\end{table}

 We can read immediately from  \Tab~\ref{tab:combin} what the constraint from EWPT on $\alpha_{\phi}^{(3)}$ means. Barring cancellations, we can exclude the scalar triplets ${\bf 3}^s_0$, ${\bf 3}^s_1$ and the gauged vector singlet ${\bf 1}^v_0$ or un-gauged vector triplet ${\bf \tilde 3}^v_0$ as tree level mediators candidates if any effect arising from $\mathcal{O}_{\partial\phi}$ or $\mathcal{O}_{\phi}^{(1)}$ is observed at the LHC. The only remaining unconstrained mediators are ${\bf 1}^s_0$ and a gauged ${\bf 3}^v_0$ or un-gauged ${\bf \tilde 1}^v_0$. They will lead to a deviation of $|\alpha_{\partial \phi}|$ from zero, which may be already seen in early stages of the LHC operation; \cf, discussion in \Sec~\ref{sec:discdev}. 
 
 \Tab~\ref{tab:combin} shows in addition that ${\bf 1}^s_0$ or ${\bf 3}^v_0$ can be easily discriminated from ${\bf \tilde 1}^v_0$ by the sign of the deviation from the Standard Model induced by $\alpha_{\partial \phi}$, which basically affects all Higgs couplings in the same way and is relatively easily testable, see \Sec~\ref{sec:discdev}. As a secondary measurement, the discovery of  $\alpha_{\phi}^{(1)} \neq 0$ may discriminate between ${\bf 1}^s_0$ and ${\bf 3}^v_0$, which is, however, a difficult measurement because it involves the relative contributions of different channels. Therefore, it may be only possible during later stages of LHC operation with increasing statistics.


\subsubsection*{Multiple mediators}

In the case when more than one mediator listed in Eq.~(\ref{equ:allmed}) is present at the high energy level, the relative contributions of the mediators, listed in \Tab~\ref{tab:combin}, add in a trivial manner. Moreover, possible interactions among two or three mediators may exist and lead to a contribution to the effective coefficients. We have explored systematically such cases. The reader can find in  \App~\ref{App:fulldecom} the relevant interactions as well as their impact on the effective coefficients. Additionally we would like to point out that under specific circumstances, such as extra symmetries or peculiar choices of high-energy couplings, some cancellations can occur allowing to evade the constraint from EWPT. These conditions can be read easily from the coefficient of $\mathcal{O}_{\phi}^{(3)}$ in \App~\ref{App:fulldecom}. 

A detailed investigation of cancellations is outside the scope of this paper. There is, however, one possibility often found in phenomenological studies, see, \eg, \Refs~\cite{Georgi:1985nv,Chanowitz:1985ug,Gunion:1989ci,Gunion:1990dt,Logan:2010en}: Two triplets are used, where the custodial symmetry requires the cancellation between  the contribution of ${\bf 3}^{s}_{0}$ with that of ${\bf 3}^{s}_{1}$ to the $T$-parameter. One can see this condition explicitly in Eq.~(\ref{equ:effresult}) 
as a cancellation condition for $\mathcal{O}_{\phi}^{(3)}$. If the model predicts $\mu_\Delta^2/m_\Delta^4 \simeq 2 |\mu_{\Delta_1}|^2/m_{\Delta_1}^4$, at tree level $\alpha_\phi^{(3)} \simeq 0$ and EWPT are avoided.  In this case, $\alpha_{\partial \phi}= 2 \mu_\Delta^2/m_\Delta^4 >0 $, which looks similar to ${\bf 1}^s_0$ or ${\bf 3}^v_0$. Most importantly, in this case $\alpha_\phi^{(1)} \simeq 2 \mu_\Delta^2/m_\Delta^4$. From \Tab~\ref{tab:combin}, we can read off that this is a unique signature since the triplet vector will lead to a negative deviation from the SM. Thus, if $\alpha_\phi^{(1)} > 0$ is found, it may point to two triplet scalars with similar masses and couplings. This measurement is not to be expected at early stages of LHC, as discussed in the earlier sections, but may be possible with high
  statistics. 

\section{Perspectives beyond LHC}
\label{sec:beyond}

 In the previous sections we have seen that the operator $\mathcal{O}_{\phi}$ does not modify the couplings of the Higgs boson to the other SM particles, thus leaving no room to detect its effects via the usual Higgs searches channels. As shown from  Eqs.~(\ref{HHH}) and~(\ref{HHHH}), this operator will contribute to the $HHH$ and $HHHH$ couplings. Yet, these interactions can only be observed via double or triple Higgs production which are difficult to measure. At LHC, only qualitative statements (such as the exclusion of a a vanishing trilinear Higgs coupling) are possible; see, \eg, \Refs~\cite{Barger:2003rs,Kanemura:2008ub,Baur:2002qd,Plehn:2005nk,Baur:2009uw}. Technology beyond LHC will be needed, such as a linear collider, CLIC, or a muon collider. For the phenomenological discussion/measurement, in particular via the triple Higgs interaction, see, \eg, \Ref~\cite{Barger:2003rs}. In addition, the $HHH$ and $HHHH$ couplings will not lead to a clean signal for $\mathcal{O}
 _\phi$, since other effective operators may contribute.
On the other hand, as one can read off from Eq.~(\ref{HZW}) 
in App.~\ref{complete-lag}, new (effective) interactions of 
the types 
$H^5$ and $H^6$ are, in principle, directly proportional to this operator. We do not study these interactions here, and we do not perform a simulation of experiments beyond LHC.  
However, we point out the theoretical implications of such measurements.

\begin{table}[t]
\centering
\begin{tabular}{|c|l|}
\hline
Mediator & Coefficient: $-3\alpha_{\phi}$  \\
\hline
\hline
${\bf 1}^s_0$ &  $\frac{1}{2} \frac{\mu_{S}^{2} \lambda_{S}}{m_{S}^{4}}
+\frac{1}{3} \frac{\mu_{S}^{3} \kappa_{S}}{m_{S}^{6}} $ \\
\hline
${\bf 2}^s_{1/2}$ & $
\frac{|\lambda_{\varphi}|^{2}}{m_{\varphi}^{2}}
+
\frac{|\tilde\lambda_{\varphi}|^{2}}{m_{\varphi}^{2}}
+
2
\frac{{\rm Re}[\lambda_{\varphi} \tilde\lambda_{\varphi}^
{*}]}{m_{\varphi}^{2}}$
\\
\hline
\hline
\end{tabular}
\caption{Coefficients of the effective operator $\mathcal{O}_{\phi}$  for the single mediator case, \ie, if only one additional mediator is present. The triplet scalars are forbidden in the single mediator case, since in this case the $T$ parameter contribution cannot be canceled.
}
\label{table:Ophisinglecase}
\end{table}

The different contributions of the mediators listed in Eq.~(\ref{equ:allmed}) to the coefficient of  $\mathcal{O}_\phi$ can be found in \App~\ref{App:fulldecom}. For simplicity we focus in the rest of this section to the cases with only one mediator; we only consider mediators unconstrained by EWPT.
 In order to describe the effects of the remaining mediators ${\bf 1}^{s}_{0}$ or ${\bf 2}^{s}_{1/2}$ on $\alpha_{\phi}$, one needs the following (minimal) Lagrangian in addition to
\equ{laglhc}, which will be testable with technology beyond LHC (``BLHC''); \cf, \App~\ref{App:fulldecom}:
\begin{eqnarray}
\mathscr{L}_{\mathrm{BLHC}}
& =& \mathscr{L}_{\mathrm{LHC}}  \nonumber \\
& + & \frac{1}{3} 
\kappa_{S} 
S^{3}
+
\frac{1}{2}
\lambda_{S}
(\phi^{\dagger} \phi) S^{2}
\nonumber \\
& + & (D_{\rho} \varphi)^{\dagger}
(D^{\rho} \varphi)
-
m_{\varphi}^{2} 
\varphi^{\dagger} \varphi
+
\lambda_{\varphi}
(\phi^{\dagger} \phi)(\phi^{\dagger} \varphi)
+
\lambda_{\varphi}^{*}
(\phi^{\dagger} \phi)(\varphi^{\dagger} \phi)
\nonumber 
\\
& + &
\tilde\lambda_{\varphi}
(\phi^{\dagger} \tau^{a} \phi)
(\phi^{\dagger} \tau^{a} \varphi)
+
\tilde\lambda_{\varphi}^{ *}
(\phi^{\dagger} \tau^{a} \phi)
(\varphi^{\dagger} \tau^{a} \phi) \, .
\label{equ:blhc}
\end{eqnarray}
After integrating out the mediators, one obtains the contributions to $\alpha_\phi$ listed in \Tab~\ref{table:Ophisinglecase}. Comparing \equ{blhc} to \equ{laglhc}, one immediately notices that 
$\alpha_{\partial\phi}$ and $\alpha_{\phi}^{(1)}$ are related to high energy couplings of the form $X\phi\phi$ where $X$ is a mediator, whereas $\alpha_{\phi}$ is sensitive to couplings of the forms  $XX\phi\phi$ and  $X\phi\phi\phi$.
For the single mediator case, we can read off from \Tab~\ref{table:Ophisinglecase} that  
$|\alpha_\phi| > 0$ can be interpreted as  ${\bf 1}^{s}_{0}$ or ${\bf 2}^{s}_{1/2}$, but it is not possible to attribute this contribution to a particular coupling.

Especially interesting is the case when ${\bf 1}^{s}_{0}$  is constrained at the LHC, which
automatically implies that $|\alpha_\phi| > 0$ is to be interpreted as a doublet scalar.
 Note that if this scalar doublet is another
Higgs, \ie, takes a vev, there may be additional modifications of EWSB
which we do not consider.\footnote{We do not assume a $\varphi^4$ term
in the Lagrangian of this scalar doublet, to avoid possible effects on
EWSB. In addition, we do not have terms such as  $ |\phi^{\dagger}
\varphi|^{2}$, since they would violate the custodial symmetry and affect
the $T$ parameter at the loop level~\cite{Barbieri:2006dq}.} This means
that the scalar doublet discussed here does not need to participate in
EWSB and may still be detected at experiments beyond LHC. Also note that
it is not surprising that it does not affect EWPT at tree level, 
as another Higgs would not either.

Cases with multiple mediators are much more complicated as can be seen from \App~\ref{App:fulldecom}, and no general conclusions can be drawn.

\section{Summary and Conclusions}
\label{sec:summary}

 The Higgs field is essential to our understanding (ignorance) of the mass mechanism for the visible world. If BSM physics is present in nature, exotic Higgs couplings may be expected in all generality.
 We have considered the impact of BSM physics in the Higgs sector
 without restriction to a particular model,  focusing on effective $d=6$
 interactions built from the Higgs field and the SM gauge fields only,
 and in particular on those operators which can be generated at tree
 level; see Eqs.~(\ref{ophidphi}) and~(\ref{ophi1ophi3}). Considering
 first each operator independently, we have computed their impact on the
 SM Lagrangian, Higgs production, and Higgs decay, working in the
 so-called $Z$-scheme in which well-measured quantities
 $G_F$, $\alpha$, and $M_Z$ are taken as inputs. 

Among the effective operators in Eqs.~(\ref{ophidphi}) and~(\ref{ophi1ophi3}), we have shown that two  are accessible at LHC: $\mathcal{O}_{\partial \phi}$, which affects the Higgs-fermion and Higgs-gauge boson couplings in the same way, and $\mathcal{O}_{\phi}^{(1)}$, which affects only the Higgs-gauge boson couplings. Another operator, $\mathcal{O}_{\phi}^{(3)}$, is already strongly constrained by electroweak precision tests, and better constraints from LHC are not expected. Finally, the detection of $\mathcal{O}_{\phi}$ interactions, requires, for instance, the measurement of the Higgs self-coupling, which needs technology beyond LHC, such as a linear collider, CLIC, or a muon collider.

First of all, we have demonstrated that contributions from $\mathcal{O}_{\phi}^{(1)}$ and  $\mathcal{O}_{\partial\phi}$ cannot  be excluded in general from LEP and Tevatron bounds, except for a small fraction of the parameter space around the Tevatron excluded  $M_H$ range, for  specific signs of the SM deviations. On the other hand, the LEP bound for the Higgs mass is relatively robust with respect to the BSM couplings analyzed here, whereas on the contrary the Tevatron bound does not hold anymore in the presence of new interactions with the Higgs sector. 

As far as the impact on LHC physics is concerned, we have demonstrated that the considered effective operators may also affect the Higgs discovery potential at the LHC. Especially at low $M_H$, an early Higgs discovery may not be possible in the presence of new physics. However, with increased luminosity, a Higgs discovered is likely in either case. The discovery of the effective interactions may, on the other hand, be harder.  While $\mathcal{O}_{\partial \phi}$ may already be established at early stages of LHC for $M_H \gtrsim 160 \, \mathrm{GeV}$,  $\mathcal{O}_{\phi}^{(1)}$ requires significantly more luminosity. A discrimination of the two operators will rely on the analysis of individual Higgs discovery channels, in particular, non-inclusive search channels such as vector boson fusion. In principle,  the measurement of the sign of the deviation from the SM can also be performed, which may help to identify the new physics. 

As one of the main results, we have performed in depth a theoretical analysis, decomposing  each effective operator in Eqs.~(\ref{ophidphi}) and~(\ref{ophi1ophi3}), to identify the Lorentz character and SM quantum numbers of all their possible heavy tree-level mediators. This allows to establish then correlations between the constraints and signals of the effective interactions discussed. 

In order to even take into account  interactions among the different mediators, 
we have simultaneously integrated out all mediators again. In conjunction with the findings above, our main results can be qualitatively summarized as follows:
\begin{description}
 \item[Early signals  LHC] (from $\mathcal{O}_{\partial \phi}$ affecting the Higgs-gauge boson and Higgs-fermion couplings) may be observable if induced by  the exchange of a singlet scalar, a gauged triplet vector, or an un-gauged singlet vector, \ie, a vector which does not interact with the covariant derivative. Through the identification of the sign of the deviation from the SM, the first two of these can be discriminated from the un-gauged singlet vector.
As another possibility, a pair of (neutral and charged) triplet scalars such that their combined impact on the $T$ parameter cancels,  may induce early signals.
 \item[Later signals at LHC] (from the extraction of $\mathcal{O}_{\phi}^{(1)}$ by the comparison of processes involving Higgs-gauge boson and Higgs-fermion couplings) may discriminate among the remaining options, including  possible pairs of triplets. 
\item[Experiments beyond LHC] may measure the $\mathcal{O}_{\phi}$ effective interaction. If no departure from the SM has been previously observed  in the Higgs interactions at the LHC, the detection of an $\mathcal{O}_{\phi}$ coupling will point to  the existence of a new  scalar doublet. This doublet scalar does not necessarily take a vev. 
\end{description}
If multiple mediators are present, the conclusions for LHC observability do not change, since all single mediator contributions add up in a trivial way. However, the possible existence of new interactions among the new mediators contributing to $\mathcal{O}_{\phi}$ prohibit a clean interpretation of this case. Note that we have only considered a singlet and a triplet vector, whereas we have also found different vectors as possible mediators if they are not required to interact with the covariant derivative (``un-gauged vector'').

We conclude that modifications of the Higgs sector may be already discovered early at the LHC, but their interpretation will require significant luminosity. We have also identified one case, a doublet scalar, which may not be testable at the LHC. These conclusions are independent of a specific model if the physics BSM couples dominantly to the Higgs sector. For example, some implications of the strongly interacting light Higgs show up as a special case of our analyis.

\subsubsection*{Acknowledgments}
We thank with Daniel Hernandez, Massimo Passera, Bohdan Grzadkowski
and Jose Ramon Espinosa for useful discussions. 
We would also like to thank UAM Madrid and W{\"u}rzburg university for their hospitality during various visits when this work has been discussed.
M.B. Gavela acknowledges CICYT support through the project FPA-2009 09017 and CAM support through the project HEPHACOS, P-ESP-00346.
 W. Winter acknowledges support by the Emmy Noether program of Deutsche
Forschungsgemeinschaft (DFG), contract no. WI 2639/2-1.


\bibliographystyle{h-elsevier}
\bibliography{references}

\newpage

\appendix

\section{Complete Lagrangian}
\label{complete-lag}

Here we show the complete Lagrangian including the effective interactions:
\begin{eqnarray}
\label{HZW}
\mathcal{L}_{H,Z,W}&=&\left[(1+\alpha_{\phi}^{(1)}\frac{v^2}{2})g_0^2\frac{v^2}{4}\right]W_{\mu}^-W^{\mu+}+\left[(1+\alpha_{\phi}^{(1)}\frac{v^2}{2}+\alpha_{\phi}^{(3)}\frac{v^2}{2})\frac{g_0^2+g_0^{'2}}{8}v^2\right] Z_{\mu}Z^{\mu}\nn\\
&+&\frac{1}{2}\partial_{\mu}H\partial^{\mu}H-\left[(1-\alpha_{\phi}^{(1)}\frac{v^2}{2}-\alpha_{\phi}^{(3)}\frac{v^2}{2}-\alpha_{\partial\phi}v^2+\frac{\alpha_{\phi}v^2}{2\lambda_0})\lambda_0 v^2\right]H^2\nn\\
&+&\left[(1+3\alpha_{\phi}^{(1)}\frac{v^2}{4}-\alpha_{\phi}^{(3)}\frac{v^2}{4}-\alpha_{\partial\phi}\frac{v^2}{2})g_0^2\frac{v}{2}\right]HW_{\mu}^-W^{\mu+}\nn\\
&+&\left[(1+3\alpha_{\phi}^{(1)}\frac{v^2}{4}+3\alpha_{\phi}^{(3)}\frac{v^2}{4}-\alpha_{\partial\phi}\frac{v^2}{2})\frac{g_0^2+g_0^{'2}}{4}v\right]HZ_{\mu}Z^{\mu}\nn\\
&+&\left[(1+5\alpha_{\phi}^{(1)}\frac{v^2}{2}-\alpha_{\phi}^{(3)}\frac{v^2}{2}-\alpha_{\partial\phi}v^2)\frac{g_0^2}{4}\right]H^2W_{\mu}^-W^{\mu+}\nn\\
&+&\left[(1+5\alpha_{\phi}^{(1)}\frac{v^2}{2}+5\alpha_{\phi}^{(3)}\frac{v^2}{2}-\alpha_{\partial\phi}v^2)\frac{g_0^2+g_0^{'2}}{8}\right]H^2Z_{\mu}Z^{\mu}\nn\\
&-&\left[\lambda_0 v (1+\frac{5}{6}\frac{v^2}{\lambda_0}\alpha_{\phi}-\frac{3}{4}\alpha_{\phi}^{(1)}v^2-\frac{3}{4}\alpha_{\phi}^{(3)}v^2-\frac{3}{2}\alpha_{\partial\phi}v^2)\right] H^3\nn\\
&-&\left[\frac{\lambda_0}{4} (1+\frac{5}{2}\frac{v^2}{\lambda_0}\alpha_{\phi}-\alpha_{\phi}^{(1)}v^2-\alpha_{\phi}^{(3)}v^2-2\alpha_{\partial\phi}v^2) \right]H^4\nn\\
&+&\left[g_0^2\frac{v}{2}\alpha_{\phi}^{(1)}\right]H^3W_{\mu}^-W^{\mu+}+\left[\frac{g_0^2+g_0^{'2}}{4}v(\alpha_{\phi}^{(1)}+\alpha_{\phi}^{(3)})\right]H^3Z_{\mu}Z^{\mu}\nn\\
&+&\left[\frac{g_0^2}{8}\alpha_{\phi}^{(1)}\right]H^4W_{\mu}^-W^{\mu+}+\left[\frac{g_0^2+g_0^{'2}}{16}(\alpha_{\phi}^{(1)}+\alpha_{\phi}^{(3)})\right]H^4Z_{\mu}Z^{\mu}\nn\\
&+&\left[\alpha_{\phi}^{(3)}\frac{v}{2}+\alpha_{\partial\phi}v\right]H\partial_{\mu}H\partial^{\mu}H+\left[\frac{\alpha_{\phi}^{(3)}}{4}+\frac{\alpha_{\partial\phi}}{2}\right]H^2\partial_{\mu}H\partial^{\mu}H\nn\\
&-&\left[\alpha_{\phi}\frac{v}{4}\right]H^5-\left[\frac{\alpha_{\phi}}{24}\right]H^6\, .
\end{eqnarray}

From this Lagrangian, one can extract the mass of the $W$ boson and the relevant Higgs couplings displayed in Eqs.~(\ref{HWW})-(\ref{HHHH}) with the SM predictions being given by
\begin{eqnarray}
&&M^2_{W_{SM}}=\frac{M_{Z}^2 }{2}(1+\sqrt{1-\frac{4\pi\alpha}{\sqrt{2}G_{F}M_{Z}^2}})\,,\\
&&\lambda_{HHH_{SM}}=\frac{M_{H}^2}{2}(\sqrt{2}G_{F})^{1/2}\,,\\
&&\lambda_{HHHH_{SM}}=\frac{M_{H}^2}{4}G_{F}\,,\\
&&\lambda_{HWW_{SM}}=M^2_{Z}(\sqrt{2}G_{F})^{1/2}(1+\sqrt{1-\frac{4\pi\alpha}{\sqrt{2}G_{F}M_{Z}^2}})\,,\\
&&\lambda_{HZZ_{SM}}=M^2_{Z}(\sqrt{2}G_{F})^{1/2}\,,\\
&&\lambda_{HHWW_{SM}}=\frac{M^2_{Z}}{2}(\sqrt{2}G_{F})(1+\sqrt{1-\frac{4\pi\alpha}{\sqrt{2}G_{F}M_{Z}^2}})\,,\\
&&\lambda_{HHZZ_{SM}}=M_{Z}^2 G_{F}\,.
\label{couplSM}
\end{eqnarray}

\section{Decay Formulae\label{appendix:Decay}}

Here details of the Higgs decay formulae involving photons are given.

\subsection{$\boldsymbol{H\rightarrow \gamma\gamma}$}

In the Standard Model, the Higgs decay into two photons is mediated  by fermion and $W$ loops. The decay width reads 
\begin{eqnarray}
\Gamma_{\rm SM}(H\rightarrow\gamma\gamma)=\frac{G_{\mu}\alpha^2
 M_H^3}{128\sqrt{2}\pi^3}\left|\sum_f N_c Q_f^2A^H_{1/2}(\tau_f)+A_1^H(\tau_W)\right|^2
\end{eqnarray}
with $\tau_i=\frac{M^2_H}{4M_i^2}$,
\begin{eqnarray}
A^H_{1/2}(\tau)&=&2[\tau+(\tau-1)f(\tau)]\tau^{-2} \, ,\\
A^H_{1}(\tau)&=&-[2\tau^2+3\tau+3(2\tau-1)f(\tau)]\tau^{-2} \, ,
\end{eqnarray}
and
\begin{equation}
f(\tau)=\left\{\begin{array}{cc}\arcsin^{2} \sqrt{\tau} & \tau\leq1\\ -\frac{1}{4}\left[\log\left(\frac{1+\sqrt{1-\tau^{-1}}}{1-\sqrt{1-\tau^{-1}}}\right)-i\pi\right] & \tau>1\end{array}\right. \, .
\end{equation}
It is reasonable to count only the top quark contribution among the fermions since the couplings of the Higgs to the fermions is proportional to the fermion mass. Hence
\begin{eqnarray}
\Gamma_{\rm SM}(H\rightarrow\gamma\gamma)=\frac{G_{\mu}\alpha^2 M_H^3}{128\sqrt{2}\pi^3}\left|\frac{4}{3}A^H_{1/2}(\tau_t)+A_1^H(\tau_W)\right|^2 \, .
\end{eqnarray}
It follows that
\begin{eqnarray}
\Gamma(H\rightarrow \gamma\gamma)=\frac{\alpha^2 M_H^3 G_\mu}{128\sqrt{2}\pi^3}\left|(1-\alpha_{\partial\phi}\frac{v^2}{2})\frac{4}{3}A^{H}_{1/2}(\tau_t)+(1+\alpha_{\phi}^{(1)}\frac{v^2}{2}-\alpha_{\partial\phi}\frac{v^2}{2})A_1^H(\tau_W)\right|^2\,,
\end{eqnarray}
and
\begin{eqnarray}
\Gamma(H\rightarrow \gamma\gamma)=\Gamma_{\mathrm{SM}}(H\rightarrow\gamma\gamma)\frac{\left|(1-\alpha_{\partial\phi}\frac{v^2}{2})\frac{4}{3}A^{H}_{1/2}(\tau_t)+(1+\alpha_{\phi}^{(1)}\frac{v^2}{2}-\alpha_{\partial\phi}\frac{v^2}{2})A_1^H(\tau_W)\right|^2}{\left|\frac{4}{3}A^{H}_{1/2}(\tau_t)+A_1^H(\tau_W)\right|^2}\, .
\end{eqnarray}


\subsection{$\boldsymbol{H\rightarrow Z\gamma}$}

As for photons, this  decay is mediated by fermion and $W$ boson loops 
\begin{eqnarray}
\Gamma_{\rm SM}(H\rightarrow\gamma Z)=\frac{G_{\mu}^2M^2_W\alpha M_H^3}{64\pi^2}(1-\frac{M_Z^2}{M_H^2})^3\left|\sum_f N_f \frac{Q_f v_f}{c}B^H_{1/2}(\tau_f,\lambda_f)+B_1^H(\tau_W,\lambda_W)\right|^2
\end{eqnarray}
with
\begin{equation}
\tau_i=\frac{4M^2_i}{M^2_H},\hspace{9mm}\lambda_i=\frac{4M_i^2}{M^2_Z},\hspace{9mm}v_f=2I^3_f-4Q_f s^2 \, ,
\end{equation}
and 
\begin{eqnarray}
B^H_{1/2}(\tau,\lambda)&=&I_1(\tau,\lambda)-I_2(\tau,\lambda) \, ,\\
B^H_1(\tau,\lambda)&=&c \left\{4(3-\frac{s^2}{c^2})I_2(\tau,\lambda)+\left[(1+\frac{2}{\tau})\frac{s^2}{c^2}-(5+\frac{2}{\tau})\right]I_1(\tau,\lambda)\right\}\,.
\end{eqnarray}
The functions $I_i$ are defined by
\begin{eqnarray}
I_1(\tau,\lambda)&=&\frac{\tau\lambda}{2(\tau-\lambda)}+\frac{\tau^2\lambda^2}{2(\tau-\lambda)^2}[f(\tau^{-1})-f(\lambda^{-1})]+\frac{\tau^2\lambda}{(\tau-\lambda)^2}[g(\tau^{-1})-g(\lambda^{-1})] \, ,\\
I_2(\tau,\lambda)&=&-\frac{\tau\lambda}{2(\tau-\lambda)}[f(\tau^{-1})-f(\lambda^{-1})] \, ,
\end{eqnarray}
with
\begin{equation}
f(\tau)=\left\{\begin{array}{cc}\arcsin^{2} \sqrt{\tau} & \tau\leq1\\ -\frac{1}{4}\left[\log\left(\frac{1+\sqrt{1-\tau^{-1}}}{1-\sqrt{1-\tau^{-1}}}\right)-i\pi\right] & \tau>1\end{array}\right. 
\end{equation}
and
\begin{equation}
g(\tau)=\left\{\begin{array}{cc}\sqrt{\tau^{-1}-1}\arcsin{\sqrt{\tau}} & \tau\geq1\\ \frac{\sqrt{1-\tau^{-1}}}{2}\left[\log\left(\frac{1+\sqrt{1+\tau^{-1}}}{1-\sqrt{1+\tau^{-1}}}\right)-i\pi\right] & \tau<1\end{array}\right. \, .
\end{equation}
Finally, one obtains
\begin{eqnarray}
\Gamma(H\rightarrow Z\gamma) & =& \Gamma_{\mathrm{SM}}(H\rightarrow Z\gamma) \times  \\
& & \frac{\left|(1-\alpha_{\partial\phi}\frac{v^2}{2})\sum_f N_f\frac{Q_f v_f}{c}B^{H}_{1/2}(\tau_f)+(1+\alpha_{\phi}^{(1)}\frac{v^2}{2}-\alpha_{\partial\phi}\frac{v^2}{2})B_1^H(\tau_W)\right|^2}{\left|\sum_f N_f\frac{Q_f v_f}{c}B^{H}_{1/2}(\tau_f)+B_1^H(\tau_W)\right|^2}\, . \nonumber
\end{eqnarray}

\section {Individual Search Channels at CMS}
\label{sec:significances}

We detail in this appendix the analysis on each search channel at CMS. Before anything, general statement can be made about the behavior of all searches channels with the different effective operators:
\begin{itemize}
\item $\mathcal{O}_{\partial\phi}$ modifies all Higgs couplings by the same factor $1-\alpha_{\partial\phi}v^2$. In consequence, all significances get enhanced (depleted) with respect to the SM ones for negative (positive) values of   
$\alpha_{\partial\phi}$. 
\item{$\mathcal{O}_{\phi}^{(1)}$}: 
Given the positive sign of the $\alpha_{\phi}^{(1)}$ contribution to the couplings, see Eqs.~(\ref{HZZm}) and~(\ref{HWWm}), the general trend expected with be an increase (decrease) with respect to the SM predictions for positive (negative) values of $\alpha_{\phi}^{(1)}$.
\end{itemize}
In the following we go back on each channel individually, giving their range of utility and bringing further detail on the significance calculation. We based our calculation on the CMS TDR~\cite{Ball:2007zza}  and also followed the recommendation of \cite{Espinosa:2010vn}.

\begin{description}
\item{ $\boldsymbol{H\rightarrow ZZ\rightarrow 4\ell}$ :} this is the most promising channel for the discovery of a Higgs boson with $M_H> 130\,\mathrm{GeV}$. 
The CMS analyses~\cite{Ball:2007zza} are based on the production of the Higgs boson through gluon and vector boson fusion. We used the Poisson significance $S_P$ defined by
\begin{equation}
\sum^{s+b-1}_{i=0}\frac{e^{-b}b^i}{i!}=\int_{-\infty}^{S_P}dx\frac{e^{-x^2/2}}{\sqrt{2\pi}},
\end{equation}
 neglecting the systematic uncertainties, that have little impact.
The branching ratio $H\rightarrow ZZ^*$ increases with $\alpha_{\phi}^{(1)}$ although it remains almost equal to the SM prediction for $M_H\geq160\,\mathrm{GeV}$. As the dominant production process -- gluon fusion -- does not depend on $\alpha_{\phi}^{(1)}$, the significance 
 turns out to be close to that for the SM, for masses in that range.

\item{ $\boldsymbol{H\rightarrow WW\rightarrow 2\ell2\nu}$ :} is the dominant process and the main discovery channel for the mass range $2M_W\leq M_H\leq 2M_Z$. The CMS analyses~\cite{Ball:2007zza} considered the gluon fusion and vector boson fusion production mechanisms. We used the $ScP2$ significance,
\begin{equation}
ScP2[s,b,\Delta b]\equiv 2(\sqrt{s+b}-\sqrt{b})\sqrt{\frac{b}{b+\Delta b^2}},
\end{equation}
 with a background systematic uncertainty $\Delta b/b$ of $10\%$.
 As in the previous channel discussed,
the significance does not evolve significantly for $M_H\geq 160\,\mathrm{GeV}$.

\item{ $\boldsymbol{H\rightarrow WW\rightarrow \ell\nu jj}$ :} for Higgs boson masses between $160\,\mathrm{GeV}$ and $180\,\mathrm{GeV}$, the $H\rightarrow ZZ^*$ gets suppressed as the $H\rightarrow WW$ channel turns on; the latter allows then to
 reconstruct the Higgs mass. The CMS analyses~\cite{Ball:2007zza} consider only the vector boson fusion process. We used the $ScL'$ significance,
 \begin{equation}
 ScL'[s,b,\Delta b]\equiv\sqrt{2[(s+b+\Delta b^2)\log(1+s/(b+\Delta b^2)-s]}
 \end{equation}
 with a background systematic uncertainty $\Delta b/b$ of $16\%$.
 Both the production mechanism and the branching ratio are rescaled by the factor $1+\alpha_{\phi}^{(1)}v^2$, hence the significance is rescaled by the square of that factor.

\item{$\boldsymbol{H\rightarrow \gamma\gamma}$ :} this is the major search channel for low Higgs masses $M_H<150\,\mathrm{GeV}$. Although the branching ratio of the Higgs into two photons is small, it is more competitive than the $H\rightarrow b\overline{b}$ channel for which the QCD background is important. The production mechanism considered by the CMS analyses~\cite{Ball:2007zza} are gluon fusion, vector boson fusion as well as Higgsstrahlung. 
We have checked that the significance increases (decreases) for positive (negative) $\alpha_{\phi}^{(1)}$, as expected.

\item{$\boldsymbol{H\rightarrow \tau\tau \rightarrow \ell+\text{\bf jets}+E_{\mathrm T}^{\mathrm{miss}}}$ :} Although not the dominant one, this channel for which the Higgs boson is produced by vector boson fusion is useful in the low mass region.  $H\rightarrow \tau\tau$ is the main decay channel next to $H\rightarrow b\overline{b}$ decay and it thus helps to improve the total significance in this mass region. We used the Poisson significance with a systematic uncertainty on the background of $7.8\%$.

\end{description}

\section{Full Decomposition of Effective Operators}
\label{App:fulldecom}

This Appendix provides the full tree-level decomposition 
of the effective operators in Eqs.~\eqref{ophidphi} and \eqref{ophi1ophi3}.
The diagrams are shown in Ref.~\cite{Arzt:1994gp}, and 
the possible mediator fields are indicated in Eq.~\eqref{equ:allmed}.
With the assumption of gauged vector mediators (and absence of the 
kinetic and mass mixing
between $\phi$ and $\varphi$),
the following  renormalizable 
interactions may be induced: 
\begin{eqnarray}
\mathcal{L}_{\text{full}}
&=&
\mathcal{L}_{S}
+
\mathcal{L}_{\Delta}
+
\mathcal{L}_{\Delta_{1}}
+
\mathcal{L}_{\varphi}
+
\mathcal{L}_{S \Delta}
+
\mathcal{L}_{S \Delta_{1}}
+
\mathcal{L}_{S \varphi}
+
\mathcal{L}_{\Delta \Delta_{1}}
+
\mathcal{L}_{\Delta \varphi}
+
\mathcal{L}_{\Delta_{1} \varphi}
\nonumber 
\\
&+&
\mathcal{L}_{V}
+
\mathcal{L}_{U}
+
\mathcal{L}_{VS},
\label{Lfull}
\end{eqnarray}
where 
\begin{eqnarray}
\mathcal{L}_{S}
&=&
\frac{1}{2} 
(\partial_{\rho} S)
(\partial^{\rho} S)
-
\frac{1}{2}
m_{S}^{2}
S^{2}
+
\mu_{S}
(\phi^{\dagger} \phi) S
+
\frac{1}{3} 
\kappa_{S} 
S^{3}
+
\frac{1}{2}
\lambda_{S}
(\phi^{\dagger} \phi) S^{2},
\\
\mathcal{L}_{\Delta}
&=&
\frac{1}{2}
(D_{\rho} \Delta)^{a}
(D^{\rho} \Delta)^{a}
-
\frac{1}{2} 
m_{\Delta}^{2}
\Delta^{a} \Delta^{a}
+
\mu_{\Delta}
(\phi^{\dagger} \tau^{a} \phi)
\Delta^{a}
+
\frac{1}{2}
\lambda_{\Delta}
(H^{\dagger} H) \Delta^{a} \Delta^{a},
\\
\mathcal{L}_{\Delta_{1}}
&=&
(D_{\rho} \Delta_{1})^{\dagger a}
(D^{\rho} \Delta_{1})^{a}
-
m_{\Delta_{1}}^{2}
\Delta_{1}^{\dagger a}
\Delta_{1}^{a}
+
\left[
\mu_{\Delta_{1}}
(\phi {i} \tau^{2} \tau^{a} \phi)
\Delta_{1}^{\dagger a}
+
{\rm h.c.}
\right]
\nonumber 
\\
&
+&
\lambda_{\Delta_{1}}
(\phi^{\dagger} \phi)
\Delta_{1}^{\dagger a}
\Delta_{1}^{a}
+
\lambda_{\Delta_{1}}^{\bf 3}
(-{i} \epsilon^{abc}) (\phi^{\dagger} \tau^{a} \phi)
\Delta_{1}^{\dagger b} \Delta_{1}^{c},
\\
\mathcal{L}_{\varphi}
&=&
(D_{\rho} \varphi)^{\dagger}
(D^{\rho} \varphi)
-
m_{\varphi}^{2} 
\varphi^{\dagger} \varphi
+
\left[
\lambda_{\varphi}
(\phi^{\dagger} \phi)(\phi^{\dagger} \varphi)
+
\tilde{\lambda}_{\varphi}
(\phi^{\dagger} \tau^{a} \phi)
(\phi^{\dagger} \tau^{a} \varphi)
+
{\rm h.c.}
\right]\,,
\\
\mathcal{L}_{S \Delta}
&=&
\frac{1}{2}
\kappa_{S \Delta}
S
\Delta^{a} \Delta^{a}
+
\lambda_{S \Delta}
(\phi^{\dagger} \tau^{a} \phi)
S \Delta^{a},
\\
\mathcal{L}_{S \Delta_{1}}
&=&
\kappa_{S \Delta_{1}}
S \Delta_{1}^{\dagger a} \Delta_{1}^{a}
+
\left[
\lambda_{S \Delta_{1}}
 (\phi {i} \tau^{2} \tau^{a} \phi) S \Delta_{1}^{\dagger a}
+
{\rm h.c.}
\right]\,,
\\
\mathcal{L}_{S \varphi}
&=&
\mu_{S \varphi} S (\phi^{\dagger} \varphi)
+
{\rm h.c.}
,
\\
\mathcal{L}_{\Delta \Delta_{1}}
&=&
\kappa_{\Delta \Delta_{1}}
(-{i} \epsilon^{abc})
\Delta_{1}^{\dagger a} \Delta_{1}^{b}
\Delta^{c}
+
\left[
\lambda_{\Delta \Delta_{1}}
(-{i} \epsilon^{abc})
(\phi {i} \tau^{2} \tau^{a} \phi)
\Delta_{1}^{\dagger b} \Delta^{c}
+
{\rm h.c.}
\right]\,,
\\
\mathcal{L}_{\Delta \varphi}
&=&
\mu_{\Delta \varphi}
\Delta^{a} (\phi^{\dagger} \tau^{a} \varphi)
+
{\rm h.c.}
,
\\
\mathcal{L}_{\Delta_{1} \varphi}
&=&
\mu_{\Delta_{1} \varphi}
\Delta_{1}^{\dagger a}
(\phi {i} \tau^{2} \tau^{a} \varphi)
+
{\rm h.c.}
,
\\
\mathcal{L}_{V}
& = &
-
\frac{1}{4} 
V_{\rho \sigma} V^{\rho \sigma}
+
\frac{1}{2} m_{V}^{2} V_{\rho} V^{\rho}
-
{i} g_{V}
V_{\rho}
[(D^{\rho} \phi)^{\dagger} \phi - \phi^{\dagger} (D^{\rho} \phi)],
\\
\mathcal{L}_{U}
&=&
- \frac{1}{2} 
U^{a}_{\rho \sigma} U^{a \rho \sigma}
+
\frac{1}{2}
m_{U}^{2} U^{a}_{\rho} U^{a \rho}
-
{i} 
\frac{g_{U}}{2}
U^{a}_{\rho}
\left[
(D^{\rho} \phi)^{\dagger} 
{\tau^{a}} \phi
-
\phi^{\dagger }
{\tau^{a}}
(D^{\rho} \phi)
\right],
\\
\mathcal{L}_{VS}
&=&
-   \mu_{VS} V_{\rho} \partial^{\rho} S
.
\end{eqnarray}
After integrating out  all mediation fields,
the following effective Lagrangian emerges 
at low energies:
\begin{eqnarray}
\mathcal{L}^{\text{eff}}_{\text{full}}
&=&
\left[
\frac{\mu_{S}^{2}}{m_{S}^{2}}
+
\frac{\mu_{\Delta}^{2}}{m_{\Delta}^{2}}
+
4 
\frac{|\mu_{\Delta_{1}}|^{2}}{m_{\Delta_{1}}^{2}}
\right]
\frac{1}{2}
(\phi^{\dagger} \phi)^{2}
\nonumber 
\\
&
+
&
\left[
\frac{\mu_{S}^{2}}{m_{S}^{4}}
+
\frac{\mu_{\Delta}^{2}}{m_{\Delta}^{4}}
+
\frac{g_{V}^{2}}{m_{V}^{2}}
+
\frac{g_{U}^{2}}{4 m_{U}^{2}}
-
\frac{\mu_{S}^{2} \mu_{VS}^{2}}{m_{S}^{4} m_{V}^{2}}
\right]
\mathcal{O}_{\partial \phi}
\nonumber 
\\
&+&
2
\left[
\frac{\mu_{\Delta}^{2}}{m_{\Delta}^{4}}
+
2
\frac{|\mu_{\Delta_{1}}|^{2}}{m_{\Delta_{1}}^{4}}
-
\frac{g_{U}^{2}}{4 m_{U}^{2}}
\right]
\mathcal{O}_{\phi}^{(1)}
-
2
\left[
\frac{\mu_{\Delta}^{2}}{m_{\Delta}^{4}}
-
2
\frac{|\mu_{\Delta_{1}}|^{2}}{m_{\Delta_{1}}^{4}}
+
\frac{g_{V}^{2}}{m_{V}^{2}}
\right]
\mathcal{O}_{\phi}^{(3)}
\nonumber
\\
&
-& 3
\Biggl[
\frac{1}{2} \frac{\mu_{S}^{2} \lambda_{S}}{m_{S}^{4}}
+
\frac{1}{3} \frac{\mu_{S}^{3} \kappa_{S}}{m_{S}^{6}}
+
\frac{1}{2} \frac{\mu_{\Delta}^{2} \lambda_{\Delta}}{m_{\Delta}^{4}}
+
2
\frac{|\mu_{\Delta_{1}}|^{2} \lambda_{\Delta_{1}}}{m_{\Delta_{1}}^{4}}
+
2
\frac{|\mu_{\Delta_{1}}|^{2} \lambda_{\Delta_{1}}^{\bf 3}}
{m_{\Delta_{1}}^{4}}
+
\frac{|\lambda_{\varphi}|^{2}}{m_{\varphi}^{2}}
+
\frac{|\tilde{\lambda}_{\varphi}|^{2}}{m_{\varphi}^{2}}
+
2
\frac{{\rm Re}[\lambda_{\varphi} \tilde{\lambda}_{\varphi}^{*}]}
{m_{\varphi}^{2}}
\nonumber 
\\
&
+
&
\frac{\mu_{S} \mu_{\Delta} \lambda_{S \Delta}}
{m_{S}^{2} m_{\Delta}^{2}}
+
\frac{1}{2} \frac{\mu_{S} \mu_{\Delta}^{2} \kappa_{S \Delta}}
{m_{S}^{2} m_{\Delta}^{4}}
+
2 \frac{\mu_{S} |\mu_{\Delta_{1}}|^{2} \kappa_{S \Delta_{1}}}
{m_{S}^{2} m_{\Delta_{1}}^{4}}
+
4 
\frac{\mu_{S} {\rm Re}[\mu_{\Delta_{1}} \lambda_{S \Delta_{1}}^{*}]}
{m_{S}^{2} m_{\Delta_{1}}^{2} }
\nonumber 
\\
&
+
&
\frac{\mu_{S}^{2} |\mu_{S \varphi}|^{2}}{m_{S}^{4} m_{\varphi}^{2}}
+
2
\frac{\mu_{S} {\rm Re}[\mu_{S \varphi} \lambda_{\varphi}^{*}]}
{m_{S}^{2} m_{\varphi}^{2} }
+
2
\frac{\mu_{S} {\rm Re}[\mu_{S \varphi} \tilde{\lambda}_{\varphi}^{*}]}
{m_{S}^{2} m_{\varphi}^{2}}
\nonumber 
\\
&
-
&
4
\frac{\mu_{\Delta} {\rm Re}[\mu_{\Delta_{1}} \lambda_{\Delta
 \Delta_{1}}^{*}]}
{m_{\Delta}^{2} m_{\Delta_{1}}^{2}}
+
2 \frac{\mu_{\Delta} |\mu_{\Delta_{1}}|^{2} \kappa_{\Delta \Delta_{1}}}
{m_{\Delta}^{2} m_{\Delta_{1}}^{4}}
+
\frac{\mu_{\Delta}^{2} |\mu_{\Delta \varphi}|^{2}}
{m_{\Delta}^{4} m_{\varphi}^{2}}
+
2
\frac{\mu_{\Delta} {\rm Re}[\mu_{\Delta \varphi} \lambda_{\varphi}^{*}]}
{m_{\Delta}^{2} m_{\varphi}^{2}}
+
2
\frac{\mu_{\Delta} {\rm Re}[\mu_{\Delta \varphi} 
\tilde{\lambda}_{\varphi}^{*}]}
{m_{\Delta}^{2} m_{\varphi}^{2}}
\nonumber 
\\
&
+
&
4 \frac{|\mu_{\Delta_{1}}|^{2} |\mu_{\Delta_{1}
 \varphi}|^{2}}{m_{\Delta_{1}}^{4} m_{\varphi}^{2}}
+
4
\frac{{\rm Re}[\mu_{\Delta_{1}}^{*} 
\mu_{\Delta_{1} \varphi} \lambda_{\varphi}^{*}]}
{m_{\Delta_{1}}^{2} m_{\varphi}^{2}}
+
4
\frac{{\rm Re}[\mu_{\Delta_{1}}^{*} 
\mu_{\Delta_{1} \varphi} \tilde{\lambda}_{\varphi}^{*}]}
{m_{\Delta_{1}}^{2} m_{\varphi}^{2}}
\nonumber 
\\
&
+
&
2
\frac{\mu_{S} \mu_{\Delta} {\rm Re}[\mu_{S \varphi} \mu_{\Delta \varphi}^{*}]}
{m_{S}^{2} m_{\Delta}^{2} m_{\varphi}^{2}}
+
4
\frac{\mu_{S} {\rm Re}[\mu_{\Delta_{1}} \mu_{S \varphi} 
\mu_{\Delta_{1} \varphi}^{*}]}
{m_{S}^{2} m_{\Delta_{1}}^{2} m_{\varphi}^{2}}
+
4
\frac{\mu_{\Delta} {\rm Re}[\mu_{\Delta_{1}} 
\mu_{\Delta \varphi} \mu_{\Delta_{1} \varphi}^{*}]}
{m_{\Delta}^{2} m_{\Delta_{1}}^{2} m_{\varphi}^{2}}
\Biggr]
\mathcal{O}_{\phi}.
\label{eq:Full-Reint-OH}
\end{eqnarray}
Only the terms which contribute to the  
effective interactions
in Eqs.~\eqref{ophidphi} and \eqref{ophi1ophi3} have been included in $\mathcal{L}_{\text{full}}$, Eq.~(\ref{eq:Full-Reint-OH}).
A putative measurement of one of these effective interactions  would give information
on the high-energy  models  
in $\mathcal{L}_{\text{full}}$, Eq.~(\ref{Lfull}), 
as long as we assume a perturbative theory.

Finally, we briefly comment on the absence of the tree-level
decomposition of the effective interactions with the 
field strength of the gauge fields (Eqs.~\eqref{ula}-\eqref{olita}),
which were proved in Ref.~\cite{Arzt:1994gp}.
In the proof, the authors assumed that the vector mediators
were gauge fields which interacted with the Higgs doublet 
through the covariant derivative (as assumed here).
This assumption played an important role in the proof.
If one allows a possibility of non-gauged vector mediators,
for example a vector doublet ${\bf 2}^{v}_{1/2}$, 
the effective interactions with field strength tensors
can also be mediated at the tree level.

\end{document}